\documentclass[12pt]{article}
\usepackage{url}
\usepackage{amsfonts}
\usepackage{amsmath,amssymb,color}
\usepackage{graphicx}
\usepackage{enumitem}
\usepackage{multirow}
\usepackage{bm}
\usepackage[natbibapa]{apacite}
\usepackage{verbatim}
\usepackage{float}
\usepackage[toc,page]{appendix}
\usepackage[onehalfspacing]{setspace}
\usepackage{setspace}
\usepackage{adjustbox}
\usepackage{rotating}
\usepackage{subcaption}
\usepackage{amsthm}
\usepackage{booktabs,dcolumn,caption}
\usepackage{array}
\usepackage{subcaption}

\usepackage[dvipsnames]{xcolor}
\usepackage{tikz}

\usepackage{xr}
\usetikzlibrary{decorations.pathreplacing,angles,quotes,fit,positioning,arrows.meta}

\newcolumntype{L}[1]{>{\raggedright\let\newline\\\arraybackslash\hspace{0pt}}p{#1}}
\newcolumntype{C}[1]{>{\centering\let\newline\\\arraybackslash\hspace{0pt}}p{#1}}
\newcolumntype{R}[1]{>{\raggedleft\let\newline\\\arraybackslash\hspace{0pt}}p{#1}}

\newcommand{\yc}[1]{{{\small \color{red}  #1}}}

\newcommand{\YY}{\mbox{$\mathbf Y$}}
\newcommand{\TT}{\mbox{$\mathbf T$}}

\newtheorem{algorithm}{Algorithm}

\newtheorem{example}{Example}

\newtheorem{proposition}{Proposition}

\title{\Large Detection of Two-Way Outliers in Multivariate Data and Application to Cheating Detection in Educational Tests}
\author{Yunxiao Chen,  Yan Lu, and Irini Moustaki\\
London School of Economics and Political Science}
\date{}

\begin{document}
\maketitle

\doublespacing

\begin{abstract}

{The paper proposes a new latent variable model for the simultaneous {(two-way)} detection of outlying individuals and items for item-response-type  data.
The proposed model is a synergy between a factor model for binary responses and continuous response times that captures normal item response behaviour and a latent class model that captures the outlying individuals and items.
A statistical decision framework is developed under the proposed model that provides compound decision rules for controlling local false discovery/nondiscovery rates of outlier detection. Statistical inference is carried out under a Bayesian framework, for which a Markov chain Monte Carlo algorithm is developed. The proposed method is applied to the detection of cheating in educational tests due to item leakage using a case study of a computer-based nonadaptive licensure assessment. 
The performance of the proposed method is  evaluated by simulation studies. 

}
\end{abstract}



\section{Introduction}\label{Sec:intro}

{Factor models \citep{bartholomew2011latent}  are widely used to analyse multivariate data, especially item-response-type data which involve individuals' responses to a set of items. For example, in educational testing, unidimensional and multidimensional  factor models, which are also known as Item Response Theory (IRT) models \citep{embretson2000item,reckase2009multidimensional}, are commonly used to model test takers' responses to test items. In these applications, a latent factor is often interpreted as the ability that the test is designed to assess.
In psychology, multidimensional factor models are typically used to describe respondents' answers to items in a psychological questionnaire \citep{wirth2007item}, where the factors are interpreted as psychological traits (e.g., personality traits).  In political science, similar models, which are often known as idea point models,  are used to describe voting behaviours \citep{bafumi2005practical}, where the factors are typically interpreted as voters' political standing.

It is often the case  that real data contain outliers among both the individuals and the items. Such outliers can lead to a substantial deviation from a carefully specified factor model which may be supported by substantive theory and historical data. These outliers often provide insights about the data  and it is thus of substantive interest to detect them. One example is Differential Item Functioning \citep[DIF;][]{wainer1993differential, millsap2012statistical}, a phenomenon that is widely observed in educational  testing, psychological measurement, as well as other related areas. It happens when a subset of items do not measure subgroups (e.g., gender, race)  in the same way. Specifically, in educational testing,  it might be the case that a subset of items are ``easier'' or ``more difficult'' for a certain subgroup than the others. In this case, the subgroup of test takers and the subset of items may be viewed as outliers, for whom a factor model that fits the rest of the data may fail.
A related, while more challenging, problem is the detection of latent DIF \citep{cho2016ncme}, for which not only the DIF items but also the group membership of individuals are not known a priori (two-way detection of outliers). One such example is the detection of cheating  in  educational  tests  due  to  item  leakage that benefits test takers through the  preknowledge of leaked items \citep{cizek2017handbook}. 
Latent DIF may also exist in educational tests due to other reasons; see \cite{cho2016ncme} for a review.
Similar problems also occur in other areas besides educational testing.  For example, in political science, it has been well recognised that roll call voting data of the United States Congress can largely be described by a liberal-conservative latent dimension with some minor deviations \citep{poole1991patterns,poole1991dimensionalizing}. Through a two-way outlier-detection formulation, i.e., by detecting outlying legislators and roll calls that do not fit the unidimensional model, one may obtain a better understanding of the patterns of roll call voting that cannot be explained by the liberal-conservative dimension.  Latent DIF may also exist in psychological measurement data in which a two-way outlier-detection formulation can facilitate the discovery of minor psychological traits and the relevant groups.

While statistical methods have been well established for the detection of DIF \citep{millsap2012statistical},  models and procedures for detecting latent DIF (two-way outliers) remain to be developed.
We therefore propose a two-way outlier {detection} model to fill the gap that adds a latent class model component  upon a factor model. The factor model component serves as a baseline model for data without outliers, and a latent class model component is used to capture the two-way outliers.  Specifically,
the proposed model imposes latent class structures among both the individuals and the items, rather than only
assuming latent classes among the individuals as in the  classical latent class analysis \citep{lazarsfeld1968latent,goodman1974exploratory,allman2009identifiability}. The proposed model is closely related to, but also substantially different from, existing statistical models and methods for the detection of outliers in multivariate data \citep[see e.g.,][]{hadi1992identifying,reiser1996analysis,mavridis2008detecting,mavridis2009forward,zhou2010stable,candes2011robust,wang2015mixture,wang2018two}

Under the proposed model, statistical decision theory is  established for the detection of two-way outliers.
Motivated by compound decision theory for multiple testing
\citep{robbins1951asymptotically,zhang2003compound,efron2001empirical,efron2004large,efron2008microarrays,efron2012large,sun2007oracle,benjamini1995controlling}, we propose the local False Discovery Rate (FDR) and local False Non-discovery Rate (FNR) as compound risk measures for the detection of two-way outliers.
Decision rules are developed based on these measures, for which optimality results are established.
The statistical inference and decision making are performed under a fully Bayesian framework, for which
a Markov chain Monte Carlo (MCMC) algorithm is developed.
Since our model involves many discrete latent variables, standard MCMC algorithms such as Gibbs and Metropolis-Hastings can suffer from slow mixing \citep[e.g.,][]{richardson1997bayesian,celeux2000computational}.
We tackle this problem by applying the parallel tempering technique \citep{geyer2011importance}.

The proposed method is applied to cheating detection based on data from the single administration of a non-adaptive test. It simultaneously detects outlying test takers and items as potential cheaters and compromised items. The proposed method uses item response data and item response time data,  which are often collected in computer-based testing, for improving outlier detection.
As shown via our real data analysis and simulation studies, incorporating response time information can improve outlier detection accuracy.
Our simulation results further suggest that the proposed model is quite robust against various forms of model misspecification, even though it
relies on some parametric assumptions that may not be satisfied perfectly in practice.

The detection of test takers who benefit from item preknowledge (cheaters) and compromised items has received much attention among quantitative researchers in education.
Specifically, \cite{mcleod2003bayesian} proposed a person-fit index for the detection of cheaters in computerised adaptive testing, under an IRT model. For non-adaptive testing, \cite{belov2013detection} proposed a person-fit index for characterising the outperformance of a student on the compromised items, assuming that
the set of compromised items is known.
Under a similar setting, \cite{sinharay2017detection} proposed likelihood-ratio and score tests for the detection of cheaters, and
\cite{segall2002item} and \cite{shu2013using} proposed IRT models for item preknowledge and developed Bayesian classification procedures.
For the detection of compromised items,
\cite{o2017detecting} and \cite{wang2020detecting} proposed methods based on data from the single administration of a non-adaptive test. These approaches require knowledge of a subset of non-compromised items to first identify a set of potential cheaters. The detection of compromised items relies on the identified cheaters in the first stage.  Under an online setting where data from multiple tests are sequentially collected,  \cite{veerkamp2000detection}, \cite{zhang2014sequential}, \cite{chen2019compound}, and \cite{chen2020item} formulated the detection of compromised items as a sequential change detection problem and proposed sequential procedures. We refer the readers to three edited volumes,
\cite{wollack2013handbook}, \cite{kingston2014test} and \cite{cizek2017handbook}, for a comprehensive review of related works. Note that most of the existing methods focus on the detection of either  cheaters or  compromised items, and often require prior information which is not always available, for example a given subset of non-compromised items.  In contrast, the proposed method can simultaneously
detect both test takers with item preknowledge and  compromised items without such prior information.




The rest of the paper is organised as follows. In Section~\ref{sec:model}, we propose a
statistical model for detecting two-way outliers in multivariate data and discuss its application to cheating detection.
Statistical decision theory is developed under a Bayesian framework in
Section~\ref{sec:decision}, and Bayesian inference procedures are given in Section~\ref{sec:infer}.
The proposed method is applied to a real dataset from a licensure
test in
Section~\ref{sec:case}.
Simulation studies are presented in Section~\ref{sec:sim}
to further evaluate the performance of the proposed method under various situations. Concluding remarks  are provided in Section~\ref{sec:remark}.
The appendix contains
the proof of a theoretical result, the details of the developed MCMC algorithm, and additional simulation results.

}

\section{Proposed Two-Way Outlier Detection Model} \label{sec:model}

\subsection{A Two-Way Outlier Detection Model for Multivariate Data}\label{subsec:model}

\begin{figure}
  \centering
  \includegraphics[scale = 1.5]{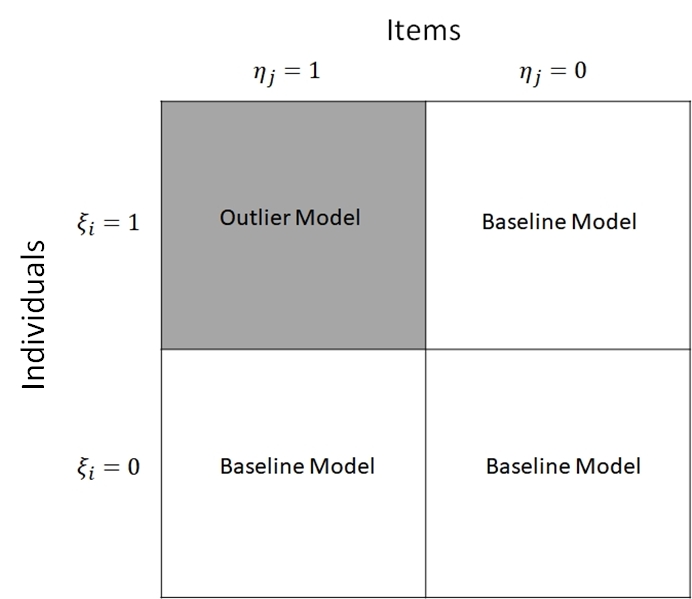}
  \caption{An illustration of the two-way outlier structure in the proposed model. }\label{fig:model}
\end{figure}

{
\subsubsection{Background and Notation}
Consider $N$ individuals responding to $J$ items. Let $Y_{ij}$ be individual $i$'s response to item $j$. We focus on binary responses, i.e., $Y_{ij}=0, 1$, where
the two types of responses may correspond to incorrect and correct answers in educational testing, and ``no" and ``yes" responses in psychological measurement, among others. We use $\YY_i = (Y_{i1}, ..., Y_{iJ})$ to denote the response vector from individual $i$ and use $\YY = (Y_{ij})_{N\times J}$ to denote the response matrix.  When item-response data are collected digitally rather than by paper and pencil, which is becoming more and more popular these days,
response time data may also be collected. Let $T_{ij}$  denote the amount of time individual $i$ spends to answer item $j$ and $\mathbf T = (T_{ij})$ denote the data matrix for response times.

In what follows, we discuss the two-way outlier detection model for  $(\YY, \TT)$. We introduce a  latent  binary variable $\xi_i$ that takes the value 1 when individual $i$ is an outlier and $0$ otherwise. Similarly, $\eta_j$ is  a latent  binary variable that takes the value 1 when item $j$ is an outlier and $0$ otherwise.
Figure~\ref{fig:model} illustrates how data are affected by the two-way outliers in the proposed model.  $(Y_{ij},T_{ij})$ are modelled with the outlier model if, and only if, both $\xi_i = 1$ and $\eta_j = 1$, which represents a typical phenomenon of latent DIF.
In the cheating detection application, items with $\eta_j = 1$ correspond to the leaked/compromised items and individuals with $\xi_i = 1$ correspond to test takers who have preknowledge about
all the compromised items before taking the test. In this context, a baseline model will capture the normal item-response behaviour, and the outlier model will capture the behaviour of the test takers with preknowledge when responding to the compromised items. In particular, the outlier model will allow test takers to have a higher probability of answering leaked items correctly and with a shorter response time \citep[see, e.g.][]{wang2018two}. The proposed model is described below.}

\subsubsection{Proposed model}{We start with a relatively more general model and then give specific examples. We introduce $\boldsymbol \theta_i$ and $\boldsymbol\tau_i$
as the person-specific parameters, also known as the factors, that drive the item responses and the response times, respectively. Both  $\boldsymbol \theta_i$ and $\boldsymbol\tau_i$ can be unidimensional or multidimensional, but their dimensions are typically assumed to be much smaller than $J$. We also introduce $\boldsymbol \beta_j$ and $\boldsymbol \alpha_j$ to denote the item-specific parameters for the item responses and the response times, respectively. To simplify the notation, we use $\Theta_i = (\boldsymbol\theta_i, \xi_i, \boldsymbol\tau_i)$ and $\Delta_j = (\boldsymbol\beta_j,  \eta_j, \boldsymbol\alpha_j)$ to denote the person- and item-specific parameter vectors, respectively.

The proposed model consists of two submodels, one for  binary  item responses and one for   continuous  response times. The item-response submodel takes the following logistic form
$$P(Y_{ij}=1 \vert \Theta_i, \Delta_j, \delta) := p(\Theta_i, \Delta_j, \delta) = \frac{\exp(h_1(\boldsymbol\theta_i, \boldsymbol\beta_j) + \xi_i\eta_j  \delta)}{1+\exp(h_1(\boldsymbol\theta_i, \boldsymbol\beta_j) + \xi_i\eta_j \delta)},$$
where $\delta$ is a non-negative parameter and $h_1(\cdot, \cdot)$ is a pre-specified function.  Given $\xi_i = 0$ or $\eta_j = 0$,
$$p(\Theta_i, \Delta_j, \delta) = \frac{\exp(h_1(\boldsymbol\theta_i, \boldsymbol\beta_j) )}{1+\exp(h_1(\boldsymbol\theta_i, \boldsymbol\beta_j) ) }$$
is the baseline item-response submodel for {non-outlying} item responses. When  $\xi_i = \eta_j = 1$, the term $\xi_i\eta_j \delta\neq 0$ captures the deviation from the baseline model. In particular, for our application, the parameter $\delta$ is set to be non-negative to let the probability of providing a correct answer (i.e., $Y_{ij} = 1$)  increase when individual $i$ and item $j$  are outliers. In this context, $\delta$ may be interpreted as the advantage that a test taker gains from item pre-knowledge. In other applications,
this sign constraint can be removed if such prior information is not available.
To keep the model parsimonious,
the parameter $\delta$ is assumed to be the same across all the outlying
individuals and items.  As will be discussed in Section~\ref{subsec:remarks}, this assumption can be relaxed.

The function $h_1$ should be chosen based on knowledge about the baseline model from substantive theory and/or historical data. We give two parametric examples of $h_1$ below, but point out that $h_1$ can also take a non-parametric form as in non-parametric IRT models \citep{ramsay1991maximum,douglas1997joint, duncan2008nonparametric}.

\begin{example} \label{exmp:rasch}
The Rasch model \citep{rasch1960probabilistic} is one of the most popular IRT models in educational testing, and is also widely used in many other areas.
In particular, the licensure test to be studied in Section \ref{sec:case} is designed and scored under this model.
The Rasch model assumes that both $\boldsymbol\theta_i$ and $\boldsymbol\beta_j$ are unidimensional. With slight abuse of notation, we denote them by non-bold typeface $\theta_i$ and $\beta_j$, respectively. This model assumes that
$h_1(\theta_i, \beta_j) = \theta_i - \beta_j$, which leads to
\begin{equation}\label{eq:ICC2}
p(\Theta_i, \Delta_j, \delta) = \frac{\exp(\theta_i - \beta_j + \xi_i\eta_j  \delta)}{1+\exp(\theta_i - \beta_j + \xi_i\eta_j \delta)}.
\end{equation}
In the context of educational testing,  $\theta_i$ and $\beta_j$ are interpreted as the ability of test taker $i$ and the difficulty of item $j$, respectively. When there are no outliers, the probability of correctly answering an item is monotone increasing with one's ability $\theta_i$ and monotone decreasing with the item's difficulty $\beta_j$.
When
$\xi_i = 1$ and $\eta_j = 1$, $\text{logit}(P(Y_{ij}=1 \vert \Theta_i, \Delta_j, \delta)) = \theta_i - \beta_j + \delta$. That is, the item response function still takes a Rasch form, but the log odds increases by a constant drift $\delta$.
This Rasch-type item-response submodel \eqref{eq:ICC2} will be further discussed in the rest of the paper, given its suitability for our case study in Section~\ref{sec:case}.
\end{example}

\begin{example}
It may be the case that the $J$ items simultaneously measure $K$ factors, $\boldsymbol\theta_i = (\theta_{i1}, ..., \theta_{iK})$, for which a multidimensional factor model is needed. In that situation, we may set
$h_1(\boldsymbol\theta_i, \boldsymbol\beta_j) = \beta_{j0} + \beta_{j1}\theta_{i1}+ \cdots + \beta_{jK}\theta_{iK},$
where $\boldsymbol\beta_j = (\beta_{j0},..., \beta_{jK})$ contains $K+1$ item-specific parameters. The item-response submodel then becomes

\begin{equation}\label{eq:ICCMIRT}
p(\Theta_i, \Delta_j, \delta)  = \frac{\exp(\beta_{j0} + \beta_{j1}\theta_{i1}+ \cdots + \beta_{jK}\theta_{iK} + \xi_i\eta_j  \delta)}{1+\exp(\beta_{j0} + \beta_{j1}\theta_{i1}+ \cdots + \beta_{jK}\theta_{iK} + \xi_i\eta_j  \delta)}.
\end{equation}
When $\xi_i = 0$ or $\eta_j = 0$, \eqref{eq:ICCMIRT} becomes the multidimensional two-parameter logistic model \citep{reckase2009multidimensional} which includes the  two-parameter logistic model \citep{birnbaum1968some} as a special case when $K = 1$.

\end{example}

The response-time submodel is specified similarly to the item-response submodel.
Specifically, we consider a log-normal model which assumes that
$$\log(T_{ij})\vert  \Theta_i, \Delta_j, \gamma, \kappa  ~~\sim~~ N\left(h_2(\boldsymbol\tau_i, \boldsymbol\alpha_j) - \xi_i\eta_j\gamma,  \kappa\right),$$
where $\gamma$ is another non-negative parameter that plays a similar role to $\delta$ in the item-response submodel, $h_2(\cdot, \cdot)$ is a pre-specified function, and $\kappa> 0$ is the variance of the normal distribution. In our application, the parameter $\gamma$ is set to be non-negative to allow the response time for outlying individuals and items to be shorter (test takers with preknowledge tend to answer the compromised items faster).
 That is, when $\xi_i = 1$ and $\eta_j = 1$, the mean log-time is reduced from the baseline level $h_2(\boldsymbol\tau_i, \boldsymbol\alpha_j)$ to $h_2(\boldsymbol\tau_i, \boldsymbol\alpha_j)  - \gamma$.
In that context, $\gamma$ may be interpreted as the reduction in response time due to item preknowledge. Similar to the discussion about parameter $\delta$, the sign constraint on $\gamma$ can also be removed if there is no such prior knowledge about the response times. We assume that the same $\gamma$ and $\kappa$ are shared by all the individuals and items
for model parsimony, which can be relaxed.

The choice of function $h_2$ is similar to the choice of function $h_1$ in the item-response submodel. In what follows, we give a specific example, but also point out that other choices of $h_2$ are possible. In particular, one can choose $h_2$ so that the baseline response-time submodel is consistent with the one proposed in \cite{van2007hierarchical}.

\begin{example}
Similar to the Rasch-type model in Example~\ref{exmp:rasch}, we let both $\boldsymbol\tau_i$ and $\boldsymbol \alpha_j$ be unidimensional and denote them by non-bold typeface $\tau_i$ and $\alpha_j$. We let function $h_2$ take the form $h_2(\tau_i, \alpha_j) = \alpha_j - \tau_i$, which leads to
\begin{equation}\label{eq:ITF2}
\log(T_{ij})\vert  \Theta_i, \Delta_j, \gamma, \kappa  ~~\sim~~ N\left(\alpha_{j} - \tau_i - \xi_i\eta_j\gamma,  \kappa\right).
\end{equation}
When $\xi_i = 0$ or $\eta_j = 0$, we obtain the baseline model for response times
$$\log(T_{ij})\vert  \Theta_i, \Delta_j, \gamma, \kappa  ~~\sim~~ N\left(\alpha_{j} - \tau_i,  \kappa\right).$$
In the context of educational testing,  $\tau_i$ can be interpreted as the speed factor of test taker $i$ and $\alpha_j$ can be interpreted as the time-consumingness of item $j$. When there are no outliers, the mean response time is monotone increasing with the item-specific time-consumingness $\alpha_j$ and  monotone decreasing with the person-specific speed factor $\tau_i$. This response-time submodel will be applied to our case study in Section~\ref{sec:case}.
\end{example}

Like many other latent variable models, conditional independence assumptions are imposed. We first assume that $(Y_{ij}, T_{ij})$, $j = 1, ..., J$, are conditionally independent given $\Theta_i$, $\Delta_j$, $\delta$, $\gamma$, and $\kappa$. Such a conditional independence assumption across items is often known as the local independence assumption. We further assume that $Y_{ij}$ and $T_{ij}$ are conditionally independent given $\Theta_i$, $\Delta_j$, $\delta$, $\gamma$, and $\kappa$, meaning that all the person effects on the response and response time distribution are captured by the person parameters. Such conditional independence assumptions are commonly made in latent variable models for item responses and response times. We refer the readers to \cite{van2007hierarchical} for the substantive justifications.

We further adopt a Bayesian hierarchical modelling framework, under which
parameters $\Theta_i$, $\Delta_j$, $\delta$, $\gamma$, and $\kappa$
are treated  as random variables.
Specifically, we let $\Theta_i$, $i = 1, ..., N$, be independent and identically distributed (i.i.d.) samples from distribution $g_1(\Theta \vert \boldsymbol \nu_1)$ and $\Delta_j, j = 1, ..., J$, be i.i.d. samples from distribution $g_2(\Delta \vert \boldsymbol \nu_2)$, respectively, where $g_1$ and $g_2$ characterise the population of individuals
and the domain of items, respectively.
Both $g_1$ and $g_2$ are taken to be parametric distributions and use $\boldsymbol \nu_1$ and $\boldsymbol \nu_2$ as generic notations for the hyperparameters of the two distributions, respectively.
This hierarchical modelling structure is visualised in Figure~\ref{fig:DAG2} using a graphical model representation.
We showcase the specification of $g_1$, $g_2$, and the priors for $\boldsymbol \nu_1, \boldsymbol \nu_2$, $\delta$, $\gamma$, and $\kappa$  in Section~\ref{subsec:priors} under the specific model with item-response submodel \eqref{eq:ICC2} and response-time submodel \eqref{eq:ITF2}.


}

\begin{figure}
  \centering
  \includegraphics[scale = 1.5]{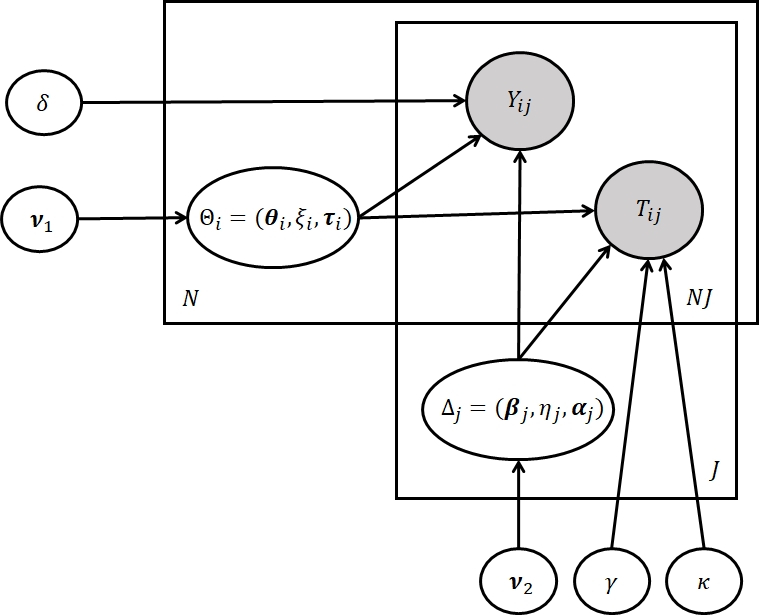}
  \caption[Path diagram 2]{Graphical representation of the proposed model for the joint distribution of item responses and response times. The boxes are plates representing replicates. The two outer plates represent individuals and items, respectively, and the inner plates present item response and response time, respectively. }\label{fig:DAG2}
\end{figure}

\subsubsection{Model without response time data} 
\begin{figure}
  \centering
  \includegraphics[scale = 1.5]{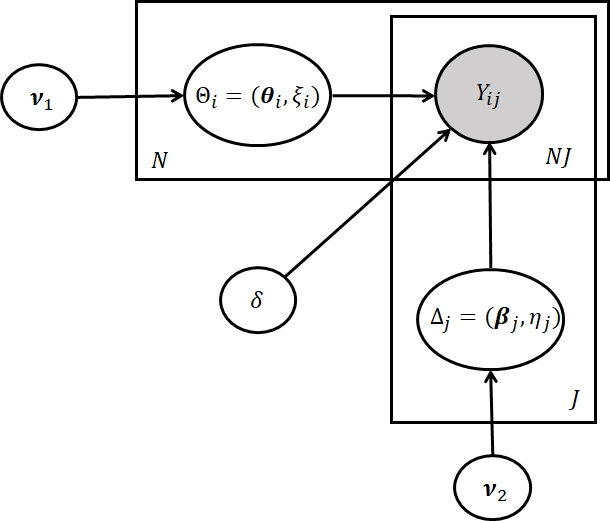}
  \caption[Path diagram 1]{Graphical representation of the reduced model when all the response times are missing completely at random. The boxes are plates representing replicates. The two outer plates represent individuals and items, respectively, and the inner plate presents an item response. }\label{fig:DAG1}
\end{figure}

Sometimes, response time information is not collected, for example, in paper-and-pencil-based educational tests.  In that case, response times are missing completely at random and the
proposed model reduces to a model for item responses.
This reduced model only contains parameters from the item-response submodel and the corresponding hyperpriors. The graphical representation of this reduced model is given in Figure~\ref{fig:DAG1}, where the reduced person
and item parameters are denoted by $\Theta_i = (\boldsymbol\theta_i, \xi_i)$ and $\Delta_j = (\boldsymbol \beta_j, \eta_j)$, respectively, and the corresponding hyperparameters  are still denoted by $\boldsymbol\nu_1$ and $\boldsymbol\nu_2$, respectively.


\subsubsection{Application to the detection of item preknowledge}\label{subsubsec:cheating}


{The proposed model framework requires that the baseline model is correctly specified. Any deviation from it is solely attributed to item preknowledge and not to other aberrant situations, such as more than one latent dimension (multidimensionality) needed to explain the associations among the items. Although this assumption might appear to be strong, it can still be examined using historical test data with no leaked items and test takers from the same population.}


Furthermore, the validity of the model interpretation also depends on the extent to which our parametric assumptions hold. Section~\ref{subsec:remarks} discusses how some of those parametric assumptions can be violated in practice and can be also relaxed.
In addition, as shown via simulation studies in Section~\ref{sec:sim},
the proposed two-way outlier detection model tends to be robust against several forms of model misspecification.

{Finally, we emphasise that,
given the sensitivity of decisions regarding cheating in tests and the relatively strong assumptions of the proposed model, the latent classes resulting from the two-way detection should be interpreted with caution (i.e. leaked items and test takers with pre-knowledge).
Results from our model can provide warnings to the test administrators, but the detected outlying cases should be further investigated and verified using additional sources of information.}

{
\subsection{Model Generalisations}\label{subsec:remarks}

Key components of the proposed two-way outlier detection model are the interaction terms  $\xi_i\eta_j  \delta$ and $\xi_i\eta_j  \gamma$ in the item-response and response-time submodels, respectively.
Specifically, the effects of the two-way outliers are
characterised by the parameters $\delta$ and $\gamma$ in the two submodels, respectively, and they are assumed to be the same across all the outliers.
This assumption can be relaxed to allow for heterogeneity among the outliers.
One way to relax this assumption is by assuming each drift parameter
to be the sum of
a person-specific parameter and an item-specific parameter. For example,
one may replace $\xi_i\eta_j  \delta$ by $\xi_i\eta_j  (\delta_i + \delta_j')$ in the item-response submodel, where $\delta_i$ and $\delta_j'$ are non-negative person- and item-specific drift parameters, respectively.

{Moreover, in the current framework, the outlier model is essentially unidimensional,
as a result of the imposed two-way latent class structure. In the application to the detection of item preknowledge, it means that a test taker has preknowledge of either  all or none of the leaked items. However, when  there are multiple sources of item leakage and test takers with item preknowledge have access to one or more of those sources, this assumptions can be relaxed by assuming multiple latent classes among both the individual and item outliers. }

}

\subsection{Related Works}

{Factor analysis in the presence of outliers has received much attention in the literature, {but mainly focuses on the detection of outlying cases/individuals rather than items as well}. One line of research is on the robust estimation of factor models \citep{moustaki2006bounded,pison2003robust,yuan1998robust,yuan2001effect}.
Another line of research focuses on the detection of outliers among the individuals who do not fit a baseline factor model, using residual-based procedures \citep{reiser1996analysis} or forward search procedures \citep{hadi1992identifying,mavridis2008detecting,mavridis2009forward}.
All these works only consider outlying individuals.
The proposed two-way outlier detection method is among the very few attempts to simultaneously classify individuals and items as outliers.


Although several models that combine factor and latent class modeling have been proposed for detecting aberrant behaviours  \citep{bolt2002item,boughton2007hybrid,goegebeur2008speeded,shu2013using,wang2015mixture,wang2018two}, none of them is about two-way classification of individuals and items.}


{Another feature of our model is that it does not require any prior knowledge about the outlying individuals and items (e.g. a subset of compromised items).}
\cite{shu2013using} proposed a Deterministic, Gated item response theory Model (DGM) for data consisting only of item responses. This model makes similar assumptions to our item-response submodel, except that
(1) the DGM assumes the known status of each item (i.e., whether each item is compromised or not), and (2) the drift parameter for cheating (i.e., $\delta$ in the current model) is assumed to be person-specific in the DGM.
{
Our model is more closely related to \cite{wang2015mixture} and \cite{wang2018two} who also assume  a mixture of log-normal distribution for response times from normal and aberrant response behaviours. Like the proposed method, these works also do not require prior knowledge about the test takers with preknowledge or the leaked items.
The main difference is that  \cite{wang2015mixture} and \cite{wang2018two} focus on identifying person-item pairs for which aberrant behaviours are involved, rather than directly classifying test takers and items. Therefore, they  allow aberrance in any person-item combination, by introducing a person-and-item specific latent variable to indicate the status of each response.  By having
person-and-item specific latent variables, the models of \cite{wang2015mixture} and \cite{wang2018two}  tend to be more flexible than the proposed model, in the sense that these models allow data to deviate from the baseline model along more directions.
Consequently, these models may be preferred when data involve multiple types of
aberrant behaviours, such as rapid guessing and cheating. On the other hand, unlike the proposed method, the
models of \cite{wang2015mixture} and \cite{wang2018two}  do not directly lead to classifications of the test takers and items, let alone quantifying the uncertainty of
the classifications. To detect test takers with preknowledge and leaked items,  follow-up analysis is needed based on  the posterior distributions of the person-and-item specific latent variables. Therefore, these methods are not as straightforward as the proposed one, if the main goal is to perform the two-way detection of individuals with preknowledge and leaked items.}

\section{Statistical Decision Theory}\label{sec:decision}
{
In what follows, we provide statistical decision theory for the detection of two-way outliers under the proposed model,
assuming the model is correctly specified.
We start with the classical Bayesian decision theory and then develop
compound decision rules for the detection of outlying individuals and items.

\subsection{Bayesian Decision Theory}\label{subsec:bayesdec}

As individuals and items are essentially mathematically exchangeable, we only discuss the Bayesian decision theory for the detection of outlying individuals.
We denote $D_i$ as the decision on individual $i$, where $D_i = 1$ means flagging the individual as an outlier and $D_i = 0$ otherwise. A false positive error happens when $D_i = 1$ and $\xi_i = 0$ and a false negative error happens when $D_i = 0$ and $\xi_i = 1$.
Decisions on the detection of outlying individuals
involve a trade-off between these two types of errors, whose importance
may be asymmetric. For example, in the application to the detection of item preknowledge, a false positive error corresponds to an innocent test taker being flagged as a cheater and a false negative error
corresponds to a cheater not being flagged. These two types of errors have substantially different consequences \citep{skorupski2017case}.

To apply Bayesian decision theory to this classification problem, we need to specify the relative cost of a false positive error, denoted by $\zeta \in (0,1)$, which further implies that the relative cost of a
false negative error is $1-\zeta$. Then the Bayes risk is defined as
\begin{equation}\label{eq:BayesRisk}
\mathcal R(D_i) := \zeta P(D_i = 1, \xi_i = 0) + (1- \zeta)P(D_i = 0, \xi_i = 1).
\end{equation}
Following the classical Bayesian decision theory  \citep[see, e.g., Chapter 2,][]{shao2003mathematical}, the optimal decision rule which minimises the Bayes risk is obtained by comparing the posterior probabilities with the relative cost $\zeta$. That is, an individual is classified as an outlier if the posterior probability is larger than $\zeta$.

This Bayesian decision rule depends on the relative cost $\zeta$. However, this parameter may not be easy to specify in practice, as the relative importance of
a false positive error is often hard to quantify. In what follows, we discuss how this parameter may be chosen adaptively based on a compound risk which is obtained by aggregating information from the entire set of individuals.

\subsection{Compound Decision for Detecting Outlying Individuals}\label{subsec:compound}

We evaluate decision making at an aggregated level for all individuals.
This involves solving $N$ decision problems simultaneously and thus is called a compound decision problem \citep{robbins1951asymptotically,zhang2003compound,sun2007oracle}.
 Given a decision rule, hypothetically, the results can be classified into four categories, as summarised in Table~\ref{tab:outcome}, where $N_{00}$, $N_{01}$, $N_{10}$, and $N_{11}$ denote the numbers of true negative, false positive, false negative, and true positive, respectively.
The quality of decisions can be quantified by two quantities. One is the False Discovery Proportion (FDP) $N_{01}/\max{\{N_{\cdot 1},1\}}$, which is the proportion of non-outliers
among the detections.
In the application to cheating detection, this gives the proportion of innocent test takers among those
who are flagged as cheaters.
The denominator is chosen so that this proportion is well-defined even when
$N_{\cdot 1} =  0$. The other quantity is the False Non-discovery Proportion (FNP) $N_{10}/\max{\{N_{\cdot 0},1\}}$, which is the proportion of outliers among the non-detections.
It is worth noting that, however, the FDP and FNP cannot be directly used
because the outliers are not directly observable.
As an alternative, we use the posterior means of the FDP and FNP, which are known as the local FDR and local  FNR, respectively. Similar measures have been proposed for solving compound decision problems in
\cite{efron2001empirical}, \cite{efron2004large, efron2008microarrays, efron2012large}, among others.
Given data and a decision rule,
the local FDR and local FNR are completely determined under the proposed model.


\begin{table}
  \centering
  \begin{tabular}{lcc|c}
    \hline
              & Not flagged as outlier & Flagged as outlier &Total \\
    \hline
    Non-outlier& $N_{00}$ & $N_{01}$& $N_{0\cdot}$ \\
    Outlier   & $N_{10}$ & $N_{11}$& $N_{1\cdot}$\\
    \hline
    Total & $N_{\cdot 0}$&$N_{\cdot 1}$ &$N$\\
    \hline
  \end{tabular}
  \caption{A summary of the outcome of detecting outlying individuals. Note that this table is hypothetical, as
  in real applications, outliers and non-outliers are directly observable. }\label{tab:outcome}
\end{table}

Suppose that the consequence of a false positive error is more severe than that of a false negative error, which may be the case for the detection of cheaters in educational testing.
Then a sensible decision criterion is to minimise the local FNR, while controlling the local FDR to be below a pre-specified threshold $\rho$.
Given the practical meaning of local FDR, the threshold $\rho$ should be much easier to specify than the relative cost in the Bayesian decision rule discussed previously. For instance, by setting $\rho = 0.01$, we approximately control the proportion of non-outliers to be below 1\% among those who are detected as outliers.

Now consider a Bayesian decision rule with a relative cost $\zeta$. We discuss how the optimal $\zeta$ is determined by the above decision criterion based on the local FDR and local FNR with a given threshold $\rho$.
For ease of exposition, we use $\mathbf Z$ as a generic notation for the data, where $\mathbf Z = \YY$ when only item responses are collected and $\mathbf Z = (\YY, \mathbf T)$ when both item responses and response times are available. Specifically, given relative cost $\zeta$, the Bayesian decision for each test taker $i$ can be written as
\begin{equation}\label{eq:decision}
D_i(\zeta) = 1_{\{P(\xi_i=1\vert \mathbf Z) > \zeta\}}.
\end{equation}
Under our fully Bayesian setting and given threshold $\zeta$, the local FDR becomes
\begin{equation}\label{eq:fdr}
\text{fdr}_\zeta(\mathbf Z) = \frac{\sum_{i=1}^N D_i(\zeta) P(\xi_i=0\vert \mathbf Z)}{\max{\{\sum_{i=1}^N D_i(\zeta),1\}}},
\end{equation}
which only depends on the posterior probabilities $P(\xi_i=1\vert \mathbf Z)$, $i = 1, ..., N$. The local  FNR  can be obtained similarly as 
$$\text{fnr}_\zeta(\mathbf Z) = \frac{\sum_{i=1}^N (1-D_i(\zeta))P(\xi_i=1\vert \mathbf Z)}{\max{\{\sum_{i=1}^N (1-D_i(\zeta)),1\}}}.$$
As summarised in Proposition~\ref{prop:fdr} below, the optimal relative cost is given by $\zeta^* = \inf \{\zeta: \text{fdr}_{\zeta}(\mathbf Z) \leq \rho\}$. That is, the decision rule given by $D_i(\zeta^*)$, $i=1, ..., N$, minimises the local FNR under the constraint that the local FDR is below $\rho$. The proof of Proposition~\ref{prop:fdr} is given in Appendix A.

\begin{proposition}\label{prop:fdr}
Given data $\mathbf Z = \YY$ or $(\YY, \mathbf T)$, the local FDR $\text{fdr}_{\zeta}(\mathbf Z)$ as a function of $\zeta$ is non-increasing and left-continuous, and the local FNR  $\text{fnr}_{\zeta}(\mathbf Z)$ is non-decreasing in $\zeta$. Thus,
\begin{equation}\label{eq:optthre}
\zeta^* = \inf \{\zeta: \text{fdr}_{\zeta}(\mathbf Z) \leq \rho\}
\end{equation}
solves the optimisation
\begin{equation}\label{eq:crit}
\min_{\zeta} \text{fnr}_\zeta(\mathbf Z), ~~s.t.~~ \text{fdr}_{\zeta}(\mathbf Z)\leq \rho.
\end{equation}
The corresponding optimal decision rule is $D_i(\zeta^*) = 1_{\{P(\xi_i=1\vert \mathbf Z) > \zeta^*\}}$.
\end{proposition}

Given posterior probabilities $P(\xi_i=1\vert \mathbf Z)$, the optimal decision is easy to obtain. The computation is described in Algorithm~\ref{alg:thre}.
\begin{algorithm}[Optimal compound decision]\label{alg:thre}
Let the posterior probabilities $P(\xi_i=1\vert \mathbf Z)$ and threshold $\rho$ for local FDR be given. The  optimal relative cost $\zeta^*$ is given by the following steps.
\begin{enumerate}
  \item Sort the posterior probabilities in an increasing order. Denote the sorted values as $p_{(1)} \leq p_{(2)} \leq \cdots \leq p_{(N)}$.
  \item Compute the cumulative means $c_{(0)} \leq c_{(1)} \leq c_{(2)} \leq \cdots \leq c_{(N)}$, where
  $$c_{(0)} = 0, \mbox{~and~} c_{(i)} = \frac{\sum_{j=1}^i p_{(j)}}{i}, i = 1, ..., N.$$
  \item Let $i^* = \max{\{i: c_{(i)} \leq \rho\}}$.
\end{enumerate}
Then  the optimal relative risk is given by $\zeta^* = p_{(i^*)}$, if $i^* > 0$, and $\zeta^* = 0$, if $i^* = 0$.
\end{algorithm}

This local FDR control procedure can be viewed as the Bayesian version of the well-known Benjamini-Hochberg (BH) procedure \citep{benjamini1995controlling} for multiple hypothesis testing. The BH procedure is designed to control the FDR, which is defined as the unconditional expectation of the FDP. Unlike the proposed procedure that is based on posterior probabilities, the BH procedure achieves the control of FDR using p-values from multiple testing. Under the proposed Bayesian framework, it seems more straightforward to control local FDR as in the proposed procedure. Also note that the FDR is automatically controlled by controlling local FDR, due to the relationship between conditional and unconditional expectations.

Another possible decision criterion is to control the posterior probability of making at least one false discovery, which corresponds to the Family-Wise Error Rate (FWER) under the frequentist setting. This FWER-type criterion exerts a more stringent control over false discovery than the proposed one by its definition. Therefore, the proposed procedure has greater power at the cost of increased rates of false positives. In this sense, the proposed local FDR control procedure is more suitable when having a large number of individuals and thus a large number of decisions need to be made simultaneously.


For certain applications, false negatives may have a more significant consequence than false positives. Then it may be more suitable to minimise the local FDR while controlling local FNR.  As the definitions of local FDR and local FNR are mathematically symmetric, the above  procedure can be easily adapted.

\subsection{Compound Decision for Detecting Outlying Items}

The compound decision theory developed above can be adapted to the detection of outlying items, based on the posterior probabilities for the item-specific binary indicators $\eta_j$.
In the application to the detection of item preknowledge,
when there is sufficient evidence suggesting that an item is compromised, it should be removed to maintain the quality of the item pool. This decision problem faces
a trade-off between the financial cost for item pool replenishment
and the need of maintaining the quality of the item pool.
For high-stake tests, test fairness is usually  the first priority and thus false negatives may have a more significant consequence than false positives. In that case, it becomes more sensible to minimise the local FDR,
under the constraint that the local FNR is below a suitable threshold (e.g., 1\%).
}

\section{Bayesian Inference}\label{sec:infer}

\subsection{Prior and Hyperprior Specification}\label{subsec:priors}

\subsubsection{Prior and hyperprior specification}
{We showcase the specification of prior and hyperprior distributions under the specific model with item-response submodel \eqref{eq:ICC2} and response-time submodel \eqref{eq:ITF2}.} We start with the specification of $g_1$, the joint distribution of $\Theta_i = (\theta_i, \xi_i, \tau_i)$.  It is assumed that $(\theta_i, \tau_i)$  follows a bivariate normal distribution $N(\mathbf 0, \Sigma),$ where $\Sigma = (\sigma_{ij})_{2\times 2}.$ Note that a person's ability and speed are typically correlated, which is why we assume a bivariate normal distribution for $(\theta_i, \tau_i)$. Similar settings are adopted in existing models for item responses and response times; see e.g., \cite{van2007hierarchical}.
We further assume that the latent indicator $\xi_i$ is independent of $(\theta_i, \tau_i)$, following a Bernoulli distribution $\mbox{Bern}(\pi_1)$. This independence assumption can be relaxed by modelling the conditional distribution of $\xi_i$ given $(\theta_i, \tau_i)$, for example, by a logistic regression model. This relaxation is left for future investigation.

We now specify $g_2$, the joint distribution of $\Delta_j = (\beta_j, \eta_j, \alpha_j)$.
Similar to that of $g_1$, we first let $(\beta_j, \alpha_j)$ follow a bivariate normal distribution $N(\boldsymbol\mu, \Omega)$, where $\boldsymbol\mu = (\mu_1, \mu_2)$ and $\Omega = (\omega_{ij})_{2\times 2}$. It is further assumed that $\eta_j$ is an independent Bernoulli random variable, $\mbox{Bern}(\pi_2)$.

It remains to specify the prior for positive parameters $\delta$, $\gamma$, $\kappa$, as well as the priors for hyperparameters  $\pi_1$, $\pi_2$, $\mu_1$, $\mu_2$,  $\Omega$, and $\Sigma$.

\begin{enumerate}
  \item As $\delta$ is positive, we assume a half-Cauchy prior distribution with scale 2.5. This is regarded a weakly informative prior, following the suggestions given in \cite{gelman2006prior}, \cite{gelman2008weakly} and \cite{polson2012half}.

  \item $\gamma$ is assumed to follow the same half-Cauchy prior distribution as $\delta$ .
  \item $\kappa$ is assumed to follow an inverse Gamma distribution, $\mbox{IG}(0.001, 0.001)$, where the shape and scale parameters are both set to 0.001. This choice follows the suggestion in Chapter 5, \cite{lunn2012bugs}.
  \item $\pi_1$ and $\pi_2$ are assumed to be i.i.d., following a
   beta distribution $\text{Beta}(2,2)$. This prior distribution can be regard as a weakly informative prior given the sample and item sizes in our application. It is suggested in \cite{agresti1998approximate} and \cite{carlin2000bayes}, Chapter 2, as the prior for a proportion parameter. We choose this distribution rather than a uniform distribution, because $\pi_1$ and $\pi_2$ are believed to not locate on the boundaries of the interval $[0, 1]$.
  \item $\mu_1$ and $\mu_2$ are assumed to be i.i.d., following a normal distribution $N(0, 5^2)$. The standard deviation 5 is chosen based on the scales of $\mu_1$ and $\mu_2$ in the current application, under which this prior may be regarded as weakly informative.
  \item $\Sigma$ and $\Omega$ are assumed to be i.i.d., following an inverse Wishart distribution where the scale matrix, $\mbox{IW}(\Psi, \nu)$, $\Psi = ((2,0)^\top, (0, 2)^\top)$,
                 and the degree of freedom $\nu = 2$.
                 This choice follows the suggestion in Chapter 6, \cite{lunn2012bugs}.
   Under this prior distribution, $\sigma_{11}$, $\sigma_{22}$, $\omega_{11}$, and $\omega_{22}$ marginally follow an inverse Gamma distribution $\mbox{IG}(1/2, 1)$.
\end{enumerate}



\subsubsection{Induced priors and hyper-priors for reduced model} For the reduced model of item response data,
the priors and hyperpriors are induced by those for the full model.
For completeness, we list the induced priors and hyperpriors below.

\begin{enumerate}
  \item For $\Theta_i = (\theta_i, \xi_i)$, $\theta_i$ and $\xi_i$ are independent, following normal distribution $N(0, \sigma_{11})$ and Bernoulli distribution $\mbox{Bern}(\pi_1)$, respectively.
  \item Similarly, for $\Delta_j = (\beta_j, \eta_j)$, $\beta_j$ and $\eta_j$ are independent, following normal distribution $N(\mu_1, \omega_{11})$ and Bernoulli distribution $\mbox{Bern}(\pi_2)$, respectively.
  \item $\delta$ follows a half-Cauchy prior distribution with scale 2.5.
  \item $\pi_1$ and $\pi_2$ are i.i.d., following a
   beta distribution $\text{Beta}(2,2)$.
  \item  $\mu_1$ follows a normal distribution $N(0, 5^2)$.
  \item $\sigma_{11}$ and $\omega_{11}$ are i.i.d. $\mbox{IG}(1/2,1)$.
\end{enumerate}

\subsection{Bayesian Inference}\label{subsec:bayes}

\subsubsection{Computation}
Statistical inference is carried out under a full Bayesian setting. An MCMC algorithm
is developed for the computation\footnote{The R code for the MCMC algorithm can be found on \url{https://github.com/YanLu-stats/OD2WIRT}.}.
This computation is non-trivial, due to the presence of many discrete variables and the interactions between them. {More specifically, the model involves person- and item-specific binary latent variables $\xi_i$ and $\eta_j$. The complexity of simulating these variables by MCMC is similar to the simulation of discrete systems like mixture models and Ising-type models. Such systems typically involve many well-separated local modes and thus suffer from the issue of
slow mixing \citep[e.g.,][]{richardson1997bayesian,celeux2000computational,katzgraber2006feedback}.
Tempering methods \citep{geyer2011importance} provide a powerful tool for exploring distributions with  many local modes.  Specifically, parallel tempering, which is also known as the Metropolis-coupled Markov chain Monte Carlo,
is chosen for the computation of the proposed model. More precisely,
parallel tempering simulates multiple MCMC chains simultaneously. A Metropolis-Hastings sampler is used for the MCMC sampling within each chain.
The target distributions of these chains are obtained by tempering the original posterior density, i.e., raising the original posterior density to different powers $T^{-1} \in [0,1]$, where  $T$ is known as the `temperature'.  The original posterior density is included by setting $T = 1$.
A chain corresponding to a higher temperature tends to have a flatter target distribution, for which the MCMC sampler is less likely to be trapped at local modes and thus has fast mixing. In contrast, when the temperature is low, the MCMC chain is more likely to be trapped and thus suffers from slow mixing. Parallel tempering improves the mixing of the low-tempered MCMC chains, by exchanging information between chains with adjacent temperatures. That is,  at each iteration, a pair of  chains with adjacent temperatures is randomly chosen and
a Metropolis-Hastings update is used to decide whether to swap their parameter states.

The use of the algorithm requires some tuning, including (a) the step sizes of random-walk Metropolis-Hastings updates within each chain, (b) the number of temperature levels, and (c) the temperature values. For (a), we follow the suggestion given by \cite{roberts2001optimal}; that is, we tune the step sizes to achieve an acceptance rate around 2.3. For (b) and (c), it is suggested to follow the theoretical guidance given in \cite{atchade2011towards}.} Further details of this algorithm are given in Appendix B.
For the implementation of the decision procedures in Section~\ref{sec:decision}, the posterior distributions of $\xi_i$ and $\eta_j$ are approximated by the posterior samples from this MCMC algorithm.

{Instead of taking a fully Bayesian setting, it is also possible to
adopt an empirical Bayes framework \citep{robbins1956empirical,casella1985introduction,efron2014two}, under which $\boldsymbol \nu_1, \boldsymbol \nu_2$, $\delta$, $\gamma$, and $\kappa$ are treated as fixed parameters and estimated by maximum likelihood estimation, while the person- and item-specific parameters $\Theta_i$ and $\Delta_j$ are treated as random variables.
However, due to the complex structure of the current model, the expectation-maximisation algorithm, which is the standard tool for empirical Bayes inference, is computationally infeasible.
A tailored stochastic optimisation algorithm needs to be developed.}

\subsubsection{Model comparison}\label{subsubsec:comparison}
{We use model comparison methods to answer the following questions that may be of substantive interest in specific applications. That is, does our item response data show evidence of the existence of outlying individuals and items? If so, does item response time information help detect these outliers?
These questions may be answered by Bayesian model comparison.

To answer the first question, we compare the proposed model for item responses with the baseline item-response model which does not contain outliers. These two models are the same, including the specification of the priors and hyperpriors, except that the hyperparameters $\pi_1$ and $\pi_2$ are set to be 0 and thus $\xi_i \eta_j \delta =  0$ for all $i$ and $j$ in the baseline model. The preference of the proposed model against the baseline model suggests the existence of the outliers.


To answer the second question, we compare the proposed model for item responses and response times with a null model for the same data. This null model is the same as the proposed model, except that the drift parameter $\gamma$ in the response-time submodel is set to be zero. When $\gamma = 0$, it means that there is no difference between the outlying and non-outlying individuals  in their response-time distributions.
If the proposed model  is preferred to the null
model, it suggests that response times contain information about the outliers
thus incorporating response time information may improve the detection of outliers.}

These models are compared by the deviance information criterion (DIC; \citealp{spiegelhalter2002bayesian}),
more specifically, a marginalised DIC in which the person- and item-specific parameters are treated as latent variables and integrated out.
In comparing two models, the one with a smaller DIC value is preferred.
 We choose the marginalised DIC, instead of the conditional DIC that
incorporates the person- and item-specific variables in the focus of the analysis, because the marginalised DIC often performs better in comparing hierarchical models \citep[e.g.,][]{quintero2018comparing}.
The marginalised DIC is computed by MCMC sampling.

While computationally convenient, the DIC suffers from several caveats, including  the lack of model selection consistency and the possibility of selecting overfitted models \citep[see][and references therein]{spiegelhalter2014deviance}. Therefore, we note that the results based on the DIC need to be interpreted with caution. In future research, other model comparison criteria will be investigated and compared with the DIC, including the Bayes factor \citep{kass1995bayes} and the Bayesian information criterion \citep[BIC;][]{schwarz1978estimating} that approximates the logarithm of the Bayes factor. The Bayes factor and BIC may be theoretically more attractive because they yield consistent model selection under suitable regularity conditions. However, their computation tends to be less straightforward than the DIC. Algorithms remain to be developed for computing them under the current modelling framework.

\section{An Application to Licensure Test}\label{sec:case}

We apply the proposed method to a dataset from a computer-based non-adaptive licensure test.
The test is designed and operated under the Rasch model, which is consistent with the proposed baseline item-response submodel.
This dataset has also been analysed for the detection of cheating in several chapters of \cite{cizek2017handbook} and journal articles including \cite{sinharay2017detection} and \cite{sinharay2017statistic}. We point out that the methods in these analyses require prior information about the items' statuses. For example, \cite{sinharay2017detection} and \cite{sinharay2017statistic} require to know the compromised items a priori and focus on the detection of cheaters. Unlike these existing analyses, we focus on the simultaneous detection of cheaters and compromised items, without requiring such prior knowledge.

The test contains 170 binary-scored items ($J = 170$), for which test takers'
item responses and response times are available.
The dataset is preprocessed by removing 12 test takers with zero response time in one or multiple items, which is believed to be data recording error.
This leads to a final dataset containing 1,624 test takers ($N = 1,624$).
The testing program flagged
41 among these 1,624 test takers as likely cheaters, through a combination of data analysis and a careful investigative process which brought in other pieces of information.
By a similar investigation process, the testing program also believed that 64 among the 170 items were compromised.
These labels will be used as partial truth for validating our data analysis results, but \textit{they are not used in our model}.
It is worth noting that these labels are not the ground truth and it is possible that there were test takers and items which ought to have been flagged but were not \citep[Chapter 1,][]{cizek2017handbook}.
It is believed that the given labels are of good quality, so that they can be used
for the evaluation of detection methods.
On the other hand,  evaluation criteria based on these labels are not perfect, due to possible labelling errors.

The purpose of this analysis is twofold. First, it is used to show the effectiveness of the proposed method, through a comparison between our results and the partial truth given by the testing program.
Second, it is used to demonstrate the use of the proposed method in real tests, which may be of interest to practitioners.

\subsection{Descriptive Analysis}\label{subsec:desc}

\begin{figure}
  \centering
\begin{subfigure}[b]{0.49\textwidth}
        \includegraphics[width=\textwidth]{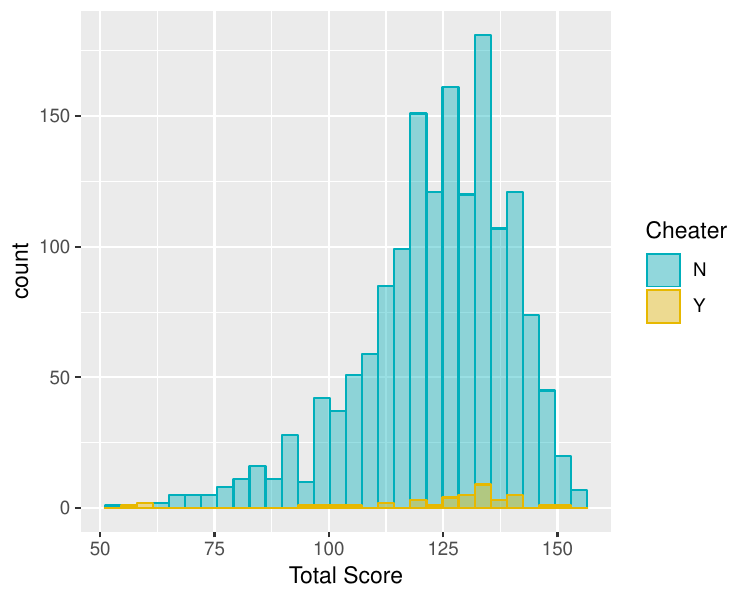}
        \caption{~~~}
    \end{subfigure}~\begin{subfigure}[b]{0.49\textwidth}
        \includegraphics[width=\textwidth]{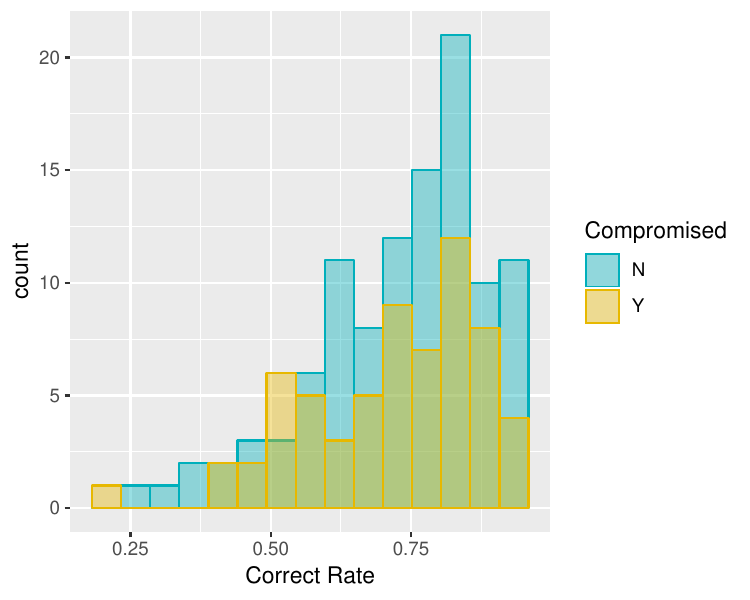}
        \caption{~~~}
    \end{subfigure}
    \caption{Descriptive analysis. Panel (a): Histogram of test takers' total scores by the testing program's cheating labels. Panel (b):
     Histogram of items' correct rates by the testing program's compromisation labels.}\label{fig:hist}
\end{figure}

\begin{figure}
  \centering
\begin{subfigure}[b]{0.49\textwidth}
        \includegraphics[width=\textwidth]{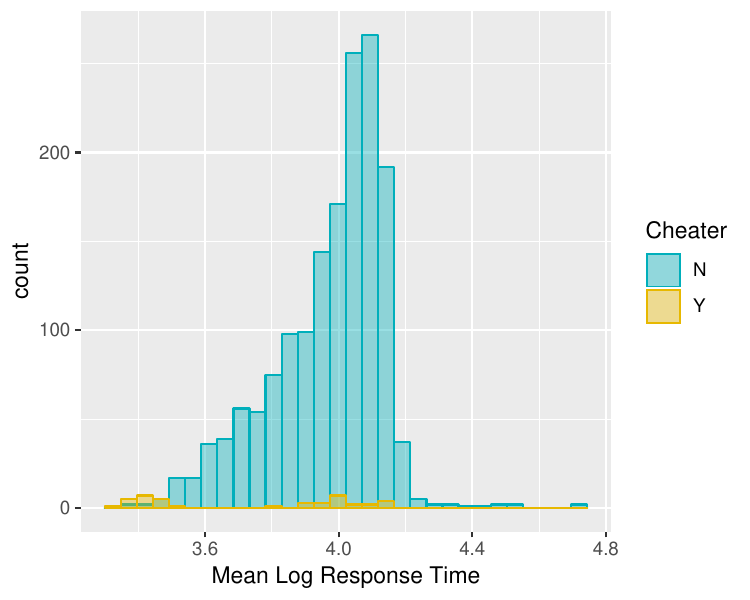}
        \caption{~~~}
    \end{subfigure}~\begin{subfigure}[b]{0.49\textwidth}
        \includegraphics[width=\textwidth]{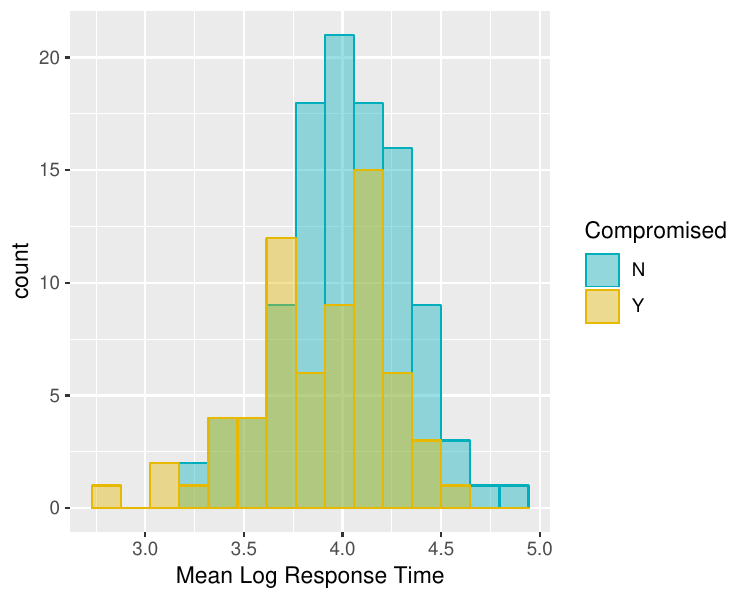}
        \caption{~~~}
    \end{subfigure}
        \caption{Descriptive analysis. Panel (a): Histogram of test takers' mean log-time by the testing program's cheating labels. Panel (b):
     Histogram of items' mean log-time by the testing program's compromisation labels.}\label{fig:hist2}
\end{figure}

We start with descriptive analysis to give an overview of the dataset. Panel (a) of Figure~\ref{fig:hist} shows the histogram of test takers' total scores by the testing program's cheating labels. Similarly, Panel (b) of Figure~\ref{fig:hist} gives the histogram of items' correct rates by the testing program's compromisation labels. Similarly, the two panels of Figure~\ref{fig:hist2} show the histograms of the mean response time in logarithm scale for test takers and items, respectively.

From these plots, it is not difficult to see that the corresponding summary statistics do not have much information about the labels on the test takers and items. In fact, the area under the curves (AUC) of the corresponding receiver operating characteristic (ROC) curves are 55.2\% and 71.7\% for the classification of the cheating labels based on total score and mean log-time, respectively.  Similarly, the corresponding AUCs for the classification of items are 52.4\% and 60.6\%, respectively. As we will see in the sequel, the proposed method substantially improves upon these benchmarks.

\subsection{Detection based on Item Responses}

We start with analysing item responses using the reduced model (i.e. only analyse item responses and not time responses). Using the algorithm given in Appendix B, three MCMC chains were run with random starting points. Their convergence was assessed by trace plots and the Gelman and Rubin (GR) diagnostic statistic \citep{gelman1992inference}.
The Gelman-Rubin R statistics applied to the parameters which are not person- or item-specific (see Table~\ref{table:est} for the list of these parameters) are below 1.20, suggesting that the chains converged to their equilibrium distributions after 10,000 iterations.

Inference is drawn based on 24,000 posterior samples from the three converged chains, where each contributes  8,000 samples.
We first compare the fitted model with its null version using the DIC measure described in Section~\ref{subsec:bayes}, to answer the question ``does our item response data show evidence of cheating?". Recall that $\pi_1 = \pi_2 = 0$ in the null model, meaning that there are no cheaters or compromised items.
The DIC value for the null model  is also based on 24,000 posterior samples from an MCMC algorithm. The DIC values for the proposed and the null models are
138,282.6 and 218,308.4, respectively. The smaller DIC for the proposed model suggests that item preknowledge is likely to exist among the test takers.

\begin{figure}
  \centering
  \begin{minipage}{0.4\textwidth}
  \includegraphics[width=0.9\textwidth]{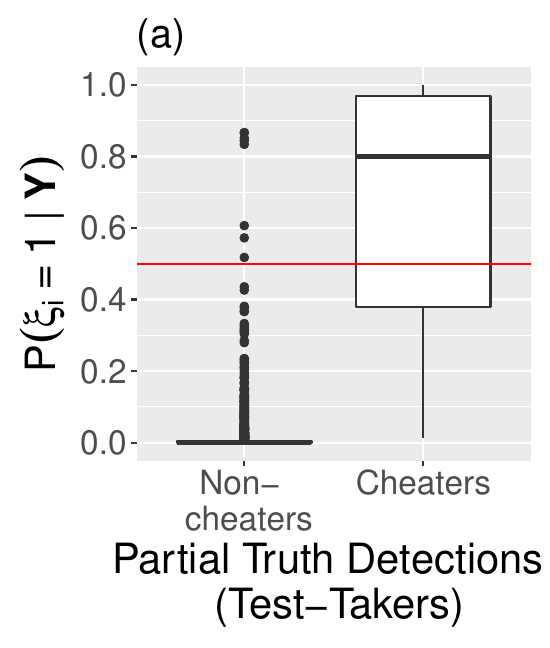}
  \end{minipage}
 \hspace{20pt}
 \begin{minipage}{0.4\textwidth}
	\includegraphics[width=0.9\textwidth]{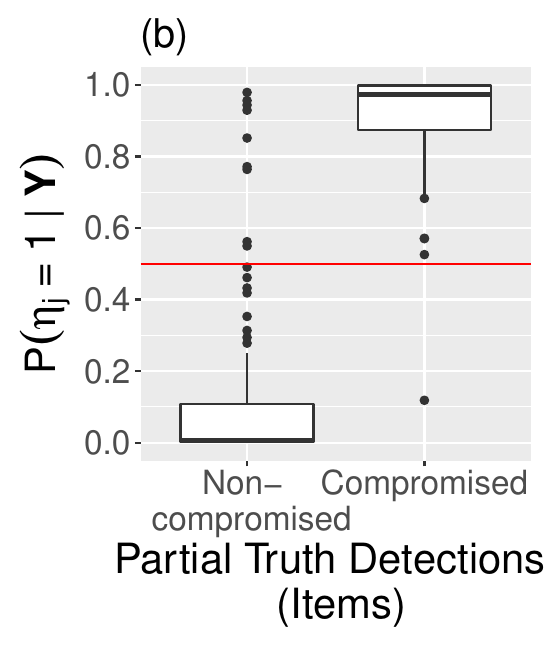}
 \end{minipage}
  \caption{Applying the reduced model for item responses. Panel (a): Box plots of the posterior means of $\xi_i$
  for the cheating and non-cheating groups (defined by the testing program). Panel (b): Box plots of the posterior means of $\eta_j$
  for the compromised and non-compromised items (defined by the testing program). 
}\label{fig:class}
\end{figure}

\begin{figure}
  \centering
  \includegraphics[width=0.7\textwidth]{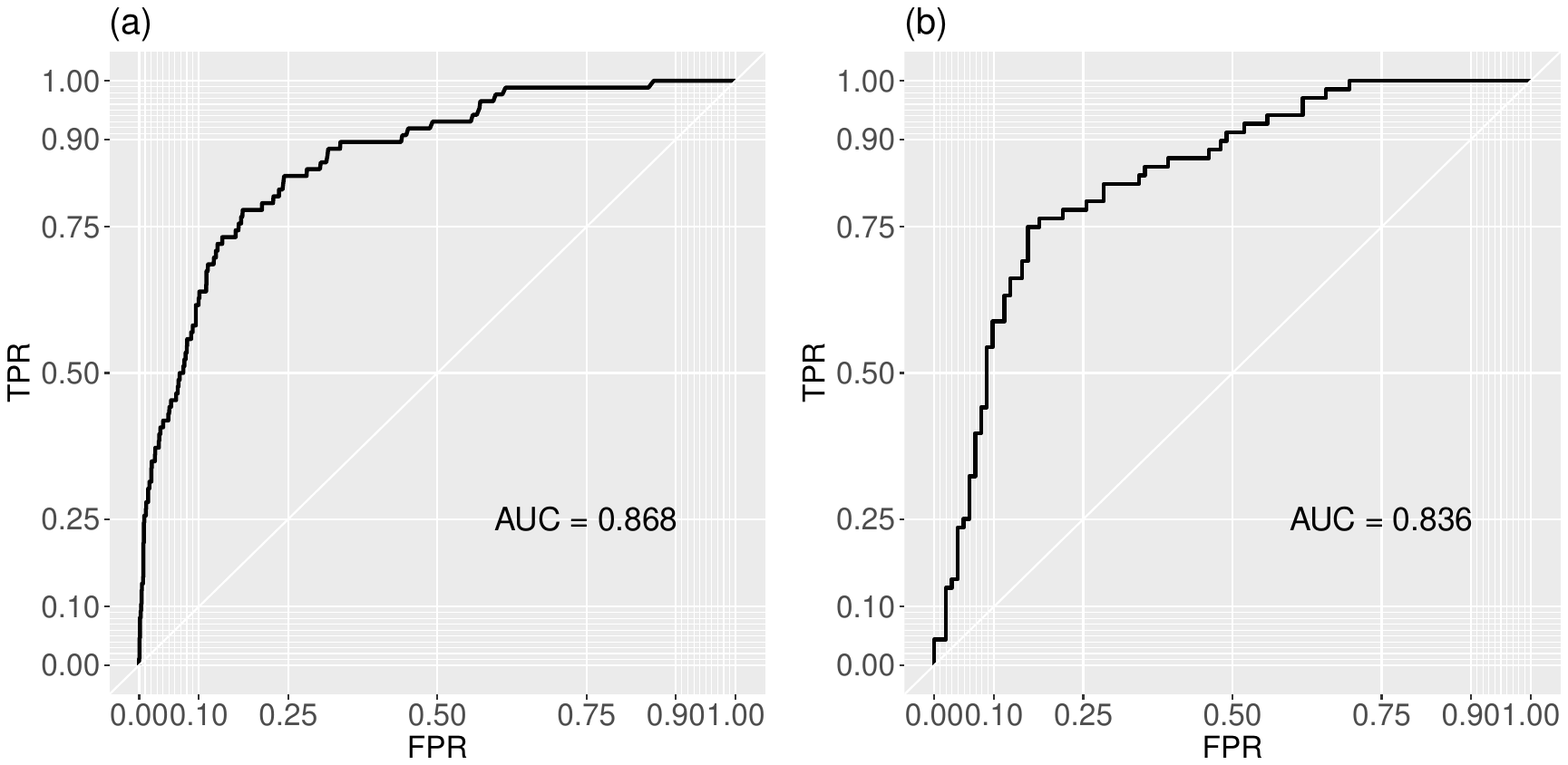}
  \caption{Applying the reduced model for item responses.  Panel (a): ROC curve for the classification of
  cheaters (labelled by the testing program) by the posterior means of $\xi_i$.
  Panel (b): ROC curve for the classification of
  compromised items (labelled by the testing program)  by
  the posterior means of $\eta_j$. The x- and y-axes of an ROC curve give the true positive rate (TPR) and false positive rate (FPR) for classification, respectively.  }\label{fig:class2}
\end{figure}

We then examine the classification results. Panel (a) of Figure~\ref{fig:class} gives the box plots of the posterior means of $\xi_i$ for the cheating and non-cheating groups (defined by the testing program), respectively.  As we can see, the posterior means of $\xi_i$ for the cheating group tend to be close to 1  and those for the non-cheating group tend to be close to 0, with some exceptions.
Panel (b) of Figure~\ref{fig:class} gives box plots of the posterior means of $\eta_j$ for the compromised and non-compromised items (defined by the testing program), respectively. Similarly, the posterior means of $\eta_j$ for the compromised items tend to be close to 1  and those for the non-compromised items tend to be close to 0.
The corresponding ROC curves for the classification of
the cheating and compromisation labels
are presented in Figure~\ref{fig:class2}.
The AUCs for these two ROC curves are 0.868 and 0.836, respectively. They are substantially larger than the ones given by the summary statistics discussed in Section~\ref{subsec:desc}. We remark that these results on the detection accuracy should be interpreted with caution, due to possible labelling errors.

\begin{figure}
  \centering
  \begin{minipage}{0.45\textwidth}
  	 \includegraphics[width=\textwidth]{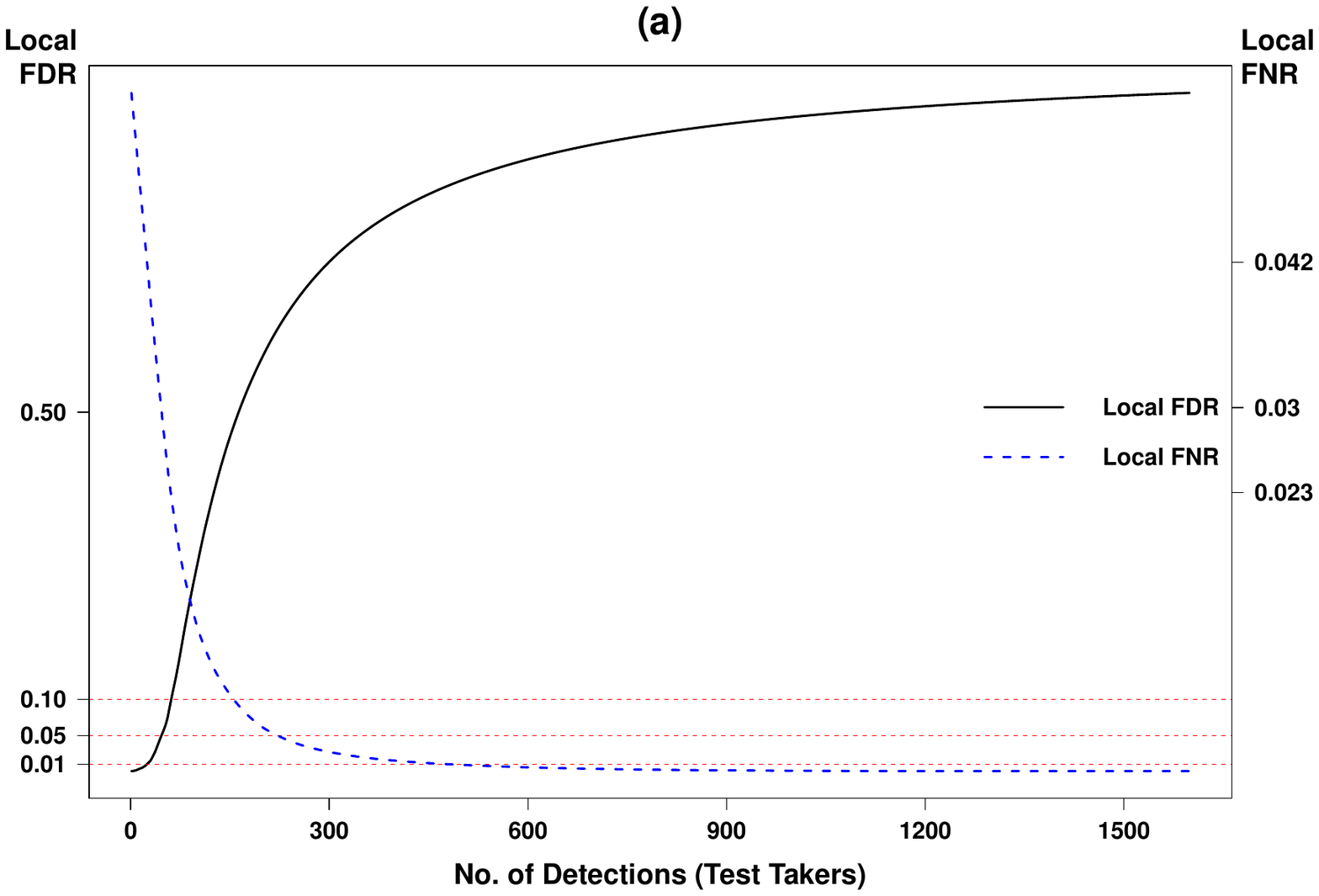}
  \end{minipage}
\begin{minipage}{0.45\textwidth}
	\includegraphics[width=\textwidth]{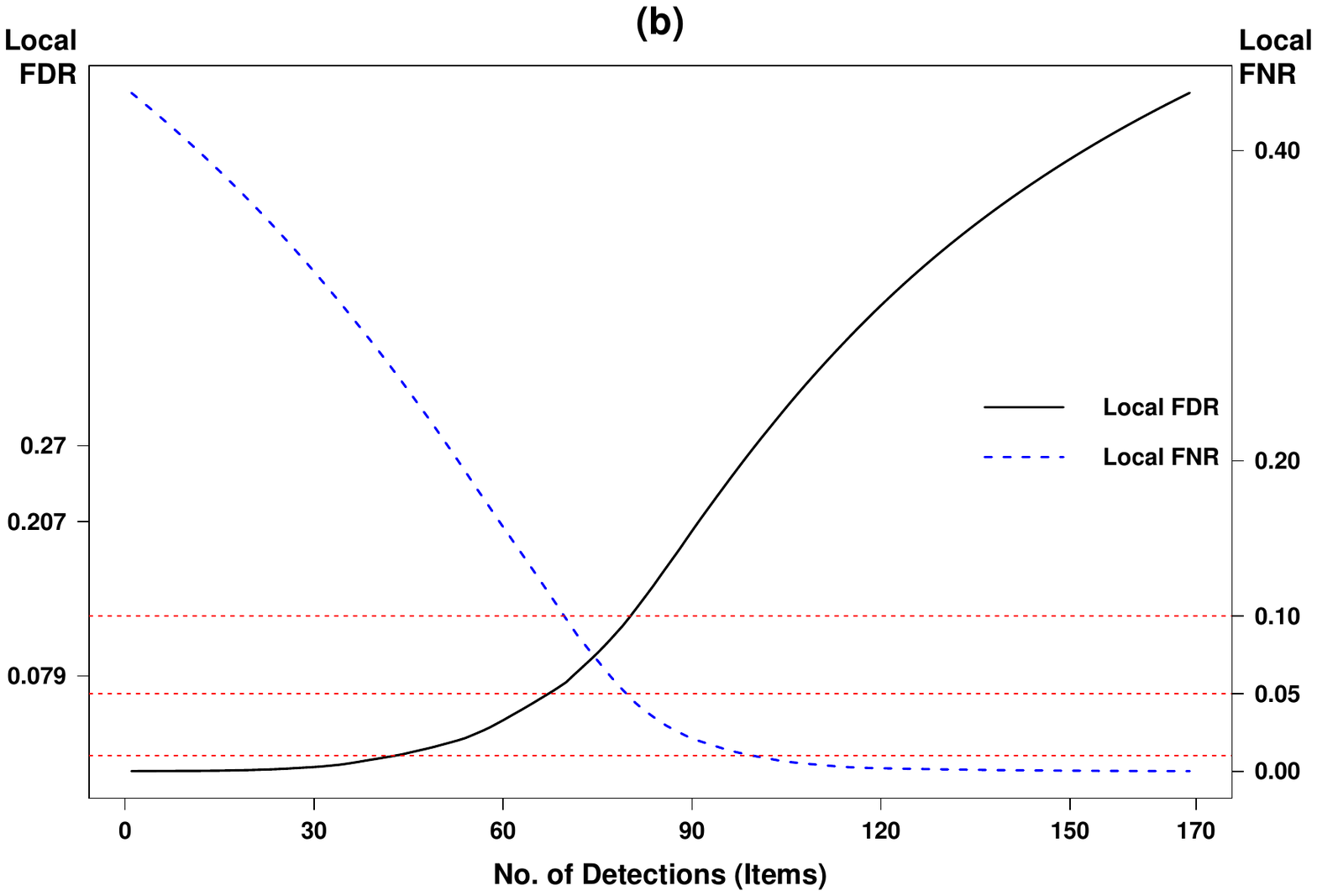}
\end{minipage}
      \caption{Applying the reduced model for item responses: The local FDR (represented by black solid curves) and the local FNR (represented by blue dashed curves) as functions of the number of detections. }\label{fig:class3}
\end{figure}

Moreover, Panel (a) of Figure~\ref{fig:class3} shows the local FDR and the local FNR as functions of the number of detections, respectively,
when applying the proposed compound decision rule to test takers.
 As we can see, as the number of detections increases, the local FDR increases and the local FNR decreases.
The same plot for items is given in Panel (b) of Figure~\ref{fig:class3}.
Specifically, the numbers of detections under different thresholds are given in Table~\ref{table:fdr}, where we control local FDR for test takers and control local FNR  for items. Again, we remark that the validity of the detection results
depends on the extent to which our model assumptions hold. Therefore, we suggest treating such detection results as  initial screening results, rather than as the final decisions.

\begin{table}
	\centering
	\begin{tabular}{l|ccc}
		\hline
		& 1\% & 5\%&10\% \\
		\hline
		Test takers & 25 & 46 & 61  \\
		Items & 100 & 91 & 71 \\
		\hline
	\end{tabular}
  \caption{Applying the reduced model for item responses.  The first row shows the numbers of detections for test takers, when controlling the corresponding local FDR at 1\%, 5\%, and 10\% levels, respectively. The second row shows the numbers of detections for items, when controlling the corresponding local FNR at 1\%, 5\%, and 10\% levels, respectively. }\label{table:fdr}
\end{table}

Finally, posterior means and 95\% credible intervals for the global parameters are presented in Table~\ref{table:est}, where the global parameters refer to the parameters that are not person-specific or item-specific.
In particular, the posterior mean of the proportion of cheaters is 2.8\%, with 95\% credible interval (2.0\%, 3.6\%). This estimate is close to the proportion of 2.5\% calculated based on the cheating labels from the testing program. The posterior mean of the proportion of compromised items is 40.1\%, with 95\% credible interval (38.7\%, 43.3\%). This estimate is close to, but slightly higher than, the proportion of 37.6\% given by the testing program. It
may be the case that
the testing program   missed several compromised items during its labelling process. Furthermore, the posterior mean of $\delta$ is 0.895. That is, the odds ratio of correctly answering a compromised item is about $\exp(0.895) = 2.447$ when comparing a cheater and a non-cheater with the same ability level.
Again, we point out that these interpretations depend on the extent to which our model holds and thus should be taken with caution. 

\begin{table}
	\centering
	\scriptsize
	\begin{tabular}{l|cccccc}
		\hline
		& $\sigma_{11}$ & $\pi_{1}$ & $\pi_{2}$ & $\omega_{11}$ & $\mu_1$ & $\delta$\\
		\hline
		EAP & $0.285$ & $0.028$ & $0.401$ & $0.685$ & $-1.004$ & $0.895$ \\
		95\% CI & $(0.261, 0.319)$ & $(0.020, 0.036)$ & $(0.387, 0.433)$ & $(0.669, 0.854)$ & $(-1.237, -0.912)$ & $(0.758, 0.959)$\\
		\hline
	\end{tabular}
  \caption{Applying the reduced model for item responses.  The row labelled ``EAP" shows the posterior means of the global parameters, where EAP represents the Expected A Posteriori, and the row labelled ``95\% CI'' provides the corresponding 95\% credible intervals. }\label{table:est}
\end{table}

\subsection{Detection based on Item Responses and Response Times}

We continue the modelling process by incorporating information from response times.
The full model is applied to the dataset consisting of both item responses and response times. Three MCMC chains were used to fit the model. According to the GR statistics applied to the global parameters, the chains converged to their equilibrium distributions after 18,000 iterations.

Similar as above, inference is drawn based on 24,000 posterior samples from the three chains after convergence. We compare this model with its null version by DIC, to answer the question ``does item response time information help detect cheating?" Recall that these two models are the same, except that the response-time drift parameter $\gamma = 0$ in the null model. The DIC values for the proposed full model and its null version are 176,935.2 and 214,201.3, respectively.
The smaller DIC value for the proposed full model suggests that response times contain substantial information about the cheating indicators.

The classification results are similar to those based only on item responses,
and thus some plots shown above are omitted here. In particular, the ROC curves based on the posterior means of $\xi_i$ and $\eta_j$ have AUCs of 0.892 and 0.867, respectively, where these AUC values are slightly higher than those from the reduced model.  In addition, the numbers of detections for test takers and items are shown in Table~\ref{table:fdr2}, where we still control local FDR for test takers and control local FNR for items. Comparing the results in Tables~\ref{table:fdr} and \ref{table:fdr2}, generally more detections tend to be made under the full model. This is likely due to the fact that the posterior distributions tend to be more concentrated under the full model as it utilises more information.

\begin{table}
	\centering
	\begin{tabular}{l|ccc}
		\hline
		& 1\% & 5\%&10\% \\
		\hline
		Test takers & 26 & 47 & 65 \\
		Items & 101 & 89 & 74 \\
		\hline
	\end{tabular}
  \caption{Applying the full model for item responses and response times. The first row shows the numbers of detections for test takers, when controlling the corresponding local FDR at 1\%, 5\%, and 10\% levels, respectively. The second row shows the numbers of detections for items, when controlling the corresponding local FNR at 1\%, 5\%, and 10\% levels, respectively. }\label{table:fdr2}
\end{table}

Posterior means and 95\% credible intervals for the global parameters are given in Table~\ref{table:est2}. Comparing Tables~\ref{table:est} and \ref{table:est2}, we find that
the estimates of the common parameters shared by the two models are close to each other. In particular, the 95\% credit intervals overlap for each parameter.
In addition, based on the posterior mean of $\Sigma$, the correlation between the ability and speed factors is as high as 0.410.
This result indicates that test takers with higher ability tend to answer the items faster.
Such a high correlation between the two factors is not uncommon for high-stake tests. For example, \cite{wang2013linear} report  a similar level of correlation between the ability and speed factors in a high-stake computerised adaptive test, under a similar Bayesian hierarchical model but without a cheating component. The estimated correlation between the two item-specific parameters is 0.237. This positive correlation suggests that solving more difficult items tends to take more time, which is consistent with our intuition.

\begin{table}
	\centering
	\scriptsize
   	\begin{tabular}{l|c c c c c c c}
	\hline
	& $\sigma_{11}$ & $\pi_{1}$ & $\pi_{2}$ & $\omega_{11}$ & $\mu_{1}$ & $\delta$\\
	\hline
	EAP & 0.289 & 0.027 & 0.410 & 0.699 & $-0.867$ & 0.807 \\
	95\% CI & (0.259, 0.298) &  (0.022,0.036) & (0.365, 0.432) & (0.626, 0.789) & $(-0.993,-0.795)$ & (0.732, 0.852) \\
	\hline
    \end{tabular}
   \vspace{10pt}

   \begin{tabular}{l|c c c c c c }
   	\hline
   	& $\sigma_{22}$ & $\sigma_{12}$ & $\omega_{22}$ & $\omega_{12}$ & $\mu_{2}$ & $\gamma$ \\
   	\hline
   	EAP &  0.248 & 0.110  & 0.397 & 0.125  & $-0.472$ & 0.620 \\
   	95\% CI & (0.213, 0.285) & (0.986, 0.139)  &  (0.334, 0.427) & (0.082, 0.132)  & $(-0.879, -0.291)$ & (0.451, 0.907) \\
   	\hline
   \end{tabular}

    \vspace{10pt}
    \begin{flushleft}
	\begin{tabular}{l|c c c }
		\hline
		 & $\kappa$\\
		\hline
		EAP   & 0.802\\
		95\% CI & (0.589, 1.037)\\
		\hline
	\end{tabular}
   \end{flushleft}
  \caption{Applying the full model for item responses and response times.  The row labelled ``EAP" shows the posterior means of the global parameters and the row labelled ``95\% CI'' provides the corresponding 95\% credible intervals.
  }\label{table:est2}
\end{table}

\section{Simulation Study}\label{sec:sim}

We now present two simulation studies for evaluating the finite-sample performance of the proposed method. The first study focuses on settings where our model is correctly specified, and the second study investigates the sensitivity of the proposed method under various forms of model misspecification.

\subsection{Study I}

\subsubsection{Settings} We consider simulation settings that mimic real educational tests. Specifically, we
consider two settings for the sample size $N$ and item size $J$, (1) $N = 2,000, J = 200$, and (2) $N = 4,000, J = 400$. This leads to two different settings, where the detection is expected to be more accurate under the second setting given its larger sample and item sizes. In what follows, these two settings are referred to as S1 and S2, respectively.

For each setting, we generate 50 independent datasets under the full model, with the
global parameters fixed across the datasets. The proportion parameters
$\pi_1$ and $\pi_2$ are set to 10\% and 40\%, respectively, the drift parameters $\delta$ and $\gamma$ are both set to be 1.2, and the rest of the
global parameters
are set to be the same as the posterior means in Table~\ref{table:est2} from the real data analysis above.
For each dataset, we apply both the reduced model for item responses and the full model for item responses and response times.
{An additional simulation study is presented in the appendix that shares a similar setting with the current study, except that the item size $J$ is set  to mimic educational tests with a smaller number of items. More specifically, this additional study considers two settings for $N$ and $J$: (1) $N = 2,000, J = 50$, and (2) $N = 4,000, J = 100$, and  the rest of the settings remain the same.  Similar results are observed in this additional study as those below from Study I.  }

\subsubsection{Results}
The analysis is conducted using our parallel tempering MCMC algorithm.
For each dataset, we run 10,000 iterations, with the first 3,000 iterations as the burn-in.  The results are based on the posterior samples from the last 7,000 iterations.

We first examine the classification results. For each model and each simulated dataset, we classify the test takers
based on the posterior means of $\xi_i$ and evaluate the classification based on the AUC value of the corresponding ROC curve. A larger AUC value implies higher classification accuracy.
Similarly, the classification of the items is based on the posterior means of $\eta_j$ and the accuracy is measured by the corresponding AUC value.
These results are  shown in Table~\ref{tab:simclass}. It can be observed that the classification is slightly more accurate under setting S2, due to the increased sample and item sizes. Moreover, the AUC values given by the full model tend to be  slightly larger than those from the reduced model, thanks to the additional information from response times.

\begin{table}
  \centering
  \footnotesize
  \begin{tabular}{ccc|cc|cc|cc}
    \hline
        & \multicolumn{4}{c|}{Test taker}  & \multicolumn{4}{c}{Item} \\
        \cline{2-9}
        & \multicolumn{2}{c|}{S1} &\multicolumn{2}{c|}{S2} & \multicolumn{2}{c|}{S1} &\multicolumn{2}{c}{S2} \\
    AUC &  Reduced & Full & Reduced & Full &Reduced & Full & Reduced & Full\\
    \hline
    25\%& 0.953 & 0.954 & 0.951 & 0.959 & 0.951 & 0.959 & 0.951 & 0.959\\
    50\%& 0.981 & 0.983 & 0.984 & 0.987 & 0.976 & 0.980 & 0.979 & 0.981\\
    75\%& 0.990 & 0.994 & 0.993 & 0.993 & 0.992 & 0.994 & 0.997 & 0.996\\
    \hline
  \end{tabular}
  \caption{Study I: Overall classification performance based on the posterior means of $\xi_i$ and $\eta_j$.
  For each model, each setting, and each target (test taker/item), we show the 25\%, 50\%, and 75\% quantiles of    the AUCs of the corresponding ROC curves from 50 independent datasets.}\label{tab:simclass}
\end{table}

We further evaluate the proposed compound decision rules. For each dataset, we control local FDR and local FNR at levels 1\%, 5\%, and 10\% for test takers  and items, respectively. We evaluate each decision rule by examining the resulting FDP and FNP; see Section~\ref{subsec:compound} for the definitions of FDP and FNP. The results are given in Tables~\ref{tab:simfdr} and \ref{tab:simfdr2} for the classifications of test takers and items, respectively.
According to these tables, the FDP is well-controlled for test takers and so is the FNP for items.

\begin{table}
  \centering
  \footnotesize
  \begin{tabular}{cccc|ccc|ccc|ccccc}
    \hline
     & \multicolumn{6}{c|}{S1}  & \multicolumn{6}{c}{S2} \\
            \cline{2-13}
    & \multicolumn{3}{c|}{Reduced}  & \multicolumn{3}{c|}{Full} & \multicolumn{3}{c|}{Reduced}  & \multicolumn{3}{c}{Full} \\
    FDP & 1\% & 5\% &10\%  & 1\% & 5\% &10\%& 1\% & 5\% &10\%  & 1\% & 5\% &10\% \\
    \hline
    25\% &  0.009 & 0.038 & 0.088 & 0.007 & 0.037 & 0.086 & 0.004 & 0.026 & 0.067 & 0.004 & 0.025 & 0.072\\
    50\% &  0.012 & 0.048 & 0.091 & 0.011 & 0.048 & 0.092 & 0.007 & 0.031 & 0.079 & 0.006 & 0.029 & 0.083\\
    75\% &  0.016 & 0.052 & 0.099 & 0.015 & 0.056 & 0.096 & 0.009 & 0.039 & 0.092 & 0.007 & 0.033 & 0.088\\
    \hline
  \end{tabular}
  \caption{Study I: Local FDR control for test takers. For each model, each setting, and each local FDR target (1\%/5\%/10\%), we show the 25\%, 50\%, and 75\% quantiles of the FDPs of the corresponding classifications from 50 independent datasets. }\label{tab:simfdr}
\end{table}

\begin{table}
  \centering
    \footnotesize
  \begin{tabular}{cccc|ccc|ccc|ccccc}
    \hline
     & \multicolumn{6}{c|}{S1}  & \multicolumn{6}{c}{S2} \\
            \cline{2-13}
    & \multicolumn{3}{c|}{Reduced}  & \multicolumn{3}{c|}{Full} & \multicolumn{3}{c|}{Reduced}  & \multicolumn{3}{c}{Full} \\
    FNP & 1\% & 5\% &10\%  & 1\% & 5\% &10\%& 1\% & 5\% &10\%  & 1\% & 5\% &10\% \\
    \hline
    25\% &  0.009 & 0.049 & 0.091 & 0.010 & 0.045 & 0.089 & 0.006 & 0.024 & 0.063 & 0.007 & 0.025 & 0.065\\
    50\% &  0.011 & 0.051 & 0.098 & 0.012 & 0.048 & 0.092 & 0.010 & 0.031 & 0.079 & 0.011 & 0.033 & 0.077\\
    75\% &  0.013 & 0.059 & 0.104 & 0.012 & 0.057 & 0.096 & 0.015 & 0.039 & 0.091 & 0.012 & 0.037 & 0.089\\
    \hline
  \end{tabular}
  \caption{Study I: Local FNR control for items. For each model, each setting, and each local FNR target (1\%/5\%/10\%), we show the 25\%, 50\%, and 75\% quantiles of the FNPs of the corresponding classifications from 50 independent datasets. }\label{tab:simfdr2}
\end{table}

Finally, we show the results on the estimation of the global parameters, as these parameters have substantive interpretations in cheating detection. Specifically, we focus on the posterior mean estimator,
for which bias and variance are estimated based on the results from 50 independent replications. These results are presented in Table~\ref{tab:simest}.
The bias, in general, tends to be close to zero for all the global parameters from  both models and both settings.
In addition, the estimation tends to be more accurate under setting S2, due to the increased sample and item sizes.

\begin{table}
\centering
\footnotesize
\begin{tabular}{lcccccc|lcccccccc}
   \hline
    S1       & \multicolumn{6}{c|}{Reduced model} &S2       & \multicolumn{6}{c}{Reduced model} \\
   \hline
           &$\pi_1$ & $\pi_2$ & $\sigma_{11}$ & $\mu_1$ & $\omega_{11}$ &$\delta$&&$\pi_1$ & $\pi_2$ & $\sigma_{11}$ & $\mu_1$ & $\omega_{11}$ &$\delta$\\
   \hline
  Bias          &0.13 &0.09 &-0.15 &-0.19 &-0.11&0.13
  &Bias       &0.09 &0.05 &0.11  &-0.08  &-0.08&0.09 \\
  Variance  &0.14&0.12&0.37&0.23&0.27&0.31
  &Variance&0.11 &0.15&0.29  &0.25   &0.21  &0.27\\
   \hline
\end{tabular}
\begin{tabular}{lccccccccccccc}
   \hline
    S1       & \multicolumn{12}{c}{Full model}& ~~~~~~~~~ \\
           \hline
           &$\pi_1$ &$\pi_2$ & $\sigma_{11}$ &$\mu_1$ & $\omega_{11}$ &$\delta$ &$\sigma_{22}$ &$\sigma_{12}$ &$\mu_2$ &$\omega_{22}$ &$\omega_{12}$ &$\kappa$ &  \\
           \hline
  Bias
&0.11 &-0.08 &0.07 &-0.24 &-0.08 &0.08 &-0.12 &-0.04 &0.07 &0.14 &0.09 &-0.16 \\
  Variance
&0.16 &0.11 &0.34 &0.19 &0.32 &0.35 &0.09 &0.07 &0.12 &0.13 &0.08 &0.77 \\
     \hline
    S2       & \multicolumn{12}{c}{Full model}& ~~~~~~~~~ \\
           \hline
           &$\pi_1$ &$\pi_2$ & $\sigma_{11}$ &$\mu_1$ & $\omega_{11}$ &$\delta$ &$\sigma_{22}$ &$\sigma_{12}$ &$\mu_2$ &$\omega_{22}$ &$\omega_{12}$ &$\kappa$ &  \\
           \hline
  Bias
&0.07 &-0.03 &0.11 &-0.18 &-0.05 &0.11 &-0.08 &-0.06 &0.09 &0.15 &0.05 &-0.11 \\
  Variance
&0.12 &0.08 &0.33 &0.21 &0.34 &0.31 &0.05 &0.09 &0.08 &0.10 &0.12 &0.63 \\
  \hline
\end{tabular}
\caption{Study I: Accuracy of the posterior mean estimator of
the global parameters. The bias and variance for the posterior mean estimator are calculated based on the 50 replications. }\label{tab:simest}
\end{table}

\subsection{Study II}\label{subsec:studyII}

\subsubsection{Settings}
In this study, we investigate the sensitivity of the proposed method under several forms of model misspecification. For the clarity of simulation settings, we focus on the misspecification of the item-response submodel. That is, we generate item-response data from a misspecified model and then apply our reduced model to classify the test takers and items. The overall classification performance, as well as the performance of the proposed compound decision rules, is evaluated.
We focus on three forms of model misspecification whose details are discussed below, including the misspecification of (1) the baseline model, (2)
the relationship between $\xi_i$ and $\theta_i$, and (3) the common drift parameter. The three forms of model misspecification, together with two settings for $N$ and $J$ as in Study I, lead to six different settings as summarised in Table~\ref{tab:simset}. For each setting, except for the misspecified part, the global parameters are set the same as those in Study I.
For each setting, 50 independent datasets are generated.

\begin{table}
  \centering
  \begin{tabular}{cccc}
    \hline
    Setting & Misspecification & $N$ & $J$\\
    \hline
    S3 & (1) & 2,000& 200 \\
    S4 & (2) & 2,000& 200\\
    S5 & (3) & 2,000& 200 \\
    \hline
    S6 & (1) & 4,000& 400 \\
    S7 & (2) & 4,000& 400\\
    S8 & (3) & 4,000& 400 \\
    \hline
  \end{tabular}
  \caption{Study II: Six simulation settings, where (1)-(3) correspond to three forms of model misspecification, including the misspecification of (1) the baseline model, (2)
the relationship between $\xi_i$ and $\theta_i$, and (3) the common drift parameter.  }\label{tab:simset}
\end{table}

We now discuss the three forms of model misspecification in detail.
For the baseline model, we replace the Rasch model by the two-parameter logistic model, an IRT model that is widely used in educational testing. That is, the following item-response submodel is assumed
$$P(Y_{ij}=1 \vert \Theta_i, \Delta_j, \alpha_j, \delta) = \frac{\exp(\alpha_j(\theta_i - \beta_j) + \xi_i\eta_j  \delta)}{1+\exp(\alpha_j(\theta_i - \beta_j) + \xi_i\eta_j \delta)},$$
where $\alpha_j$ is known as the discrimination parameter. Note that the proposed model can be viewed as a special case where $\alpha_j = 1$ for all $j$. In the misspecified model, we generate the discrimination parameters $\alpha_j$ from a uniform distribution $U[1, 1.5]$.

In the proposed model, $\theta_i$ and $\xi_i$ are assumed to be independent, meaning that whether a test taker cheats or not is independent of his/her ability. This assumption may not hold and it is likely that these two variables are negatively associated, i.e., test takers with lower ability are more likely to cheat in a test.
To mimic this situation, we generate $(\theta_i,\xi_i)$ jointly from a  Gaussian copula. That is, we first generate $(\theta_i,\xi_i^*)$ from a bi-variate normal distribution, with mean vector $(-0.867, 0)^\top$ and covariance matrix
$((0.289, -0.134)^\top,(-0.134, 1)^\top)$.
Under this bivariate normal distribution, the correlation between $\theta_i$ and $\xi_i^*$ is
$-0.25$. We then let $\xi_i = 1_{\{\xi_i^* \geq z_{0.9}\}}$, which is obtained by truncating $\xi_i^*$ at $z_{0.9}$, the 90\% quantile of the standard normal distribution, so that $P(\xi_i = 1) = 0.1$. Under this Gaussian copula model, the marginal distributions of $\theta_i$
and $\xi_i$ remain the same as those in Study I, while a negative association is introduced between the two variables.

For model parsimony, it is also assumed in the proposed model that the drift parameter $\delta$ is common across all the test takers and items. This assumption may not hold in practice. Therefore, in this misspecified model, instead of using a constant drift, we assume the drift parameter to be both item- and person-specific. That is, we assume
$$P(Y_{ij}=1 \vert \Theta_i, \Delta_j, \delta_{ij})  = \frac{\exp(\theta_i - \beta_j + \xi_i\eta_j  \delta_{ij})}{1+\exp(\theta_i - \beta_j + \xi_i\eta_j \delta_{ij})},$$
where the drift parameters $\delta_{ij}$ are generated i.i.d. from a uniform distribution $U[1, 1.5]$.

\subsubsection{Results} We evaluate the proposed method under the six settings above.  Similar to Study I, we evaluate the overall classification performance by AUC and the performance of the compound decision rules by the corresponding FDP and FNP. The results are given in Tables~\ref{tab:sim2class} through \ref{tab:sim2fdr2}.
Specifically, the AUC values in Table~\ref{tab:sim2class}  are comparable to those from the correctly specified model in Table \ref{tab:simclass},
though the AUCs from settings S3, S4, S6, and S7 are slightly smaller. As further shown in Tables~\ref{tab:sim2fdr} and \ref{tab:sim2fdr2},
the compound decision rules tend to control the corresponding FDP and FNP under the targeted levels, expect when the target level is 1\%. That is, when controlling the local FDR and local FNR to be below 1\% for test takers and items, respectively, the resulting FDP and FNP tend to exceed the targeted level under all six settings. This is likely due to the fact that the posterior probabilities cannot be accurately obtained when they are close to 0 or 1, under model misspecification.

Overall, the proposed method is reasonably robust against several forms of model specification, though the performance may be slightly affected. However, under potential model misspecification, the method should be used with caution if we aim to control local FDR or local FNR to be below
  a very small threshold (e.g., 1\%). 

\begin{table}
  \centering
  \footnotesize
  \begin{tabular}{ccccccc|ccccccccccccc}
    \hline
        & \multicolumn{6}{c|}{Test takers}  & \multicolumn{6}{c}{Item} \\
        \cline{2-13}
    AUC & $S_3$ & $S_4$ & $S_5$ & $S_6$ & $S_7$ & $S_8$& $S_3$ & $S_4$ & $S_5$ & $S_6$ & $S_7$ & $S_8$\\
    \hline
    25\%& 0.904 & 0.889 & 0.945 & 0.912 & 0.922 & 0.947
    & 0.921 & 0.903 & 0.949 & 0.947 & 0.937 & 0.945\\

    50\%& 0.965 & 0.927 & 0.975& 0.969 & 0.950 & 0.981
    & 0.965 & 0.936 & 0.964 & 0.962 & 0.951 & 0.976\\

    75\%& 0.983 & 0.989 & 0.994 & 0.980 & 0.984 & 0.993
    &0.981 & 0.971 & 0.988 & 0.985 & 0.984 & 0.991\\
    \hline
  \end{tabular}
  \caption{Study II: Overall classification performance based on the posterior means of $\xi_i$ and $\eta_j$.
  For each model, each setting, and each target (test taker/item), we show the 25\%, 50\%, and 75\% quantiles of the AUCs of the corresponding ROC curves from 50 independent datasets.}\label{tab:sim2class}
\end{table}

\begin{table}
  \centering
    \footnotesize
  \begin{tabular}{cccc|ccc|ccc}
    \hline
     & \multicolumn{3}{c|}{S3}  & \multicolumn{3}{c|}{S4}& \multicolumn{3}{c}{S5}  \\
            \cline{2-10}
    FDP & 1\% & 5\% &10\%  & 1\% & 5\% &10\%& 1\% & 5\% &10\% \\
    \hline
    25\% &0.013 &0.031 &0.061 &0.018 &0.042 &0.059 &0.012 &0.028 &0.067\\
    50\% &0.015 &0.039 &0.072 &0.023 &0.048 &0.065 &0.026 &0.039 &0.072\\
    75\% &0.019 &0.046 &0.074 &0.026 &0.059 &0.070 &0.029 &0.043 &0.079\\
    \hline
         & \multicolumn{3}{c|}{S6}  & \multicolumn{3}{c|}{S7}& \multicolumn{3}{c}{S8}  \\
             \cline{2-10}
    FDP & 1\% & 5\% &10\%  & 1\% & 5\% &10\%& 1\% & 5\% &10\% \\
    \hline
    25\% &0.004 &0.029 &0.075 &0.014 &0.036 &0.059 &0.009 &0.025 &0.061\\
    50\% &0.007 &0.041 &0.089 &0.017 &0.043&0.066 &0.014 & 0.041 &0.064\\
    75\% &0.012 &0.045 &0.093 &0.024 &0.047 &0.072 &0.023 &0.052 &0.075\\
    \hline
  \end{tabular}
  \caption{Study II: Local FDR control for test takers. For each model, each setting, and each local FDR target (1\%/5\%/10\%), we show the 25\%, 50\%, and 75\% quantiles of the FDPs of the corresponding classifications from 50 independent datasets. }\label{tab:sim2fdr}
\end{table}

\begin{table}
  \centering
    \footnotesize
  \begin{tabular}{cccc|ccc|ccc}
    \hline
     & \multicolumn{3}{c|}{S3}  & \multicolumn{3}{c|}{S4}& \multicolumn{3}{c}{S5}  \\
            \cline{2-10}
    FNP & 1\% & 5\% &10\%  & 1\% & 5\% &10\%& 1\% & 5\% &10\% \\
    \hline
    25\% &0.007 &0.021 &0.061 &0.019 &0.032 &0.078 &0.015 &0.039 &0.072\\
    50\% &0.014 &0.032 &0.074 &0.022 &0.037 &0.073 &0.028 &0.043 &0.081\\
    75\% &0.017 &0.038 &0.076 &0.024 &0.041 &0.085 &0.030&0.047&0.087\\
    \hline
         & \multicolumn{3}{c|}{S6}  & \multicolumn{3}{c|}{S7}& \multicolumn{3}{c}{S8}  \\
             \cline{2-10}
    FNP & 1\% & 5\% &10\%  & 1\% & 5\% &10\%& 1\% & 5\% &10\% \\
    \hline
    25\% &0.009 &0.024 &0.045 &0.012 &0.031 &0.072 &0.014 &0.036 &0.064\\
    50\% &0.011 &0.025 &0.059 &0.015 &0.036 &0.079 &0.026 &0.046 &0.073\\
    75\% &0.014 &0.032 &0.082 &0.025 &0.041 &0.086 &0.029 &0.054 &0.083\\
    \hline
  \end{tabular}
  \caption{Study II: local FNR control for items. For each model, each setting, and each local FNR target (1\%/5\%/10\%), we show the 25\%, 50\%, and 75\% quantiles of the FNPs of the corresponding classifications from 50 independent datasets. }\label{tab:sim2fdr2}
\end{table}

\section{Concluding Remarks}\label{sec:remark}

{In this paper, we propose a Bayesian hierarchical model for the detection of latent differential item functioning in  item-response-type multivariate data. The proposed method simultaneously detects outlying individuals and items that deviate from a given baseline model. Furthermore, a compound decision theory is proposed for the detection of two-way outliers
under a Bayesian decision framework. Statistical inference is carried out under a fully Bayesian framework for which a parallel tempering
MCMC algorithm is developed.
The proposed model is largely motivated by, and applied to, the detection of test takers who benefit from item preknowledge and compromised items in educational tests.
The proposed method requires little prior knowledge about either the test takers with item preknowledge or the compromised items, and thus is directly applicable to operational tests as a monitoring tool {or more generally about outlying cases and items in other applications}.}

The proposed method is successfully applied to data from a licensure test which is known to suffer from item preknowledge.
In this study, two models are applied, including the
reduced model for item responses and the full model for item responses and response times.
Both models accurately detect the potential cheaters and compromised items identified by the testing program, suggesting their usefulness in practice. In addition, the full model performs slightly better than the reduced one, suggesting that response-time information may help detect cheating.
However, it should be noted that the labels provided by the testing program in this example are not the ground truth and thus the accuracy measures may be compromised.
The validity of the proposed method remains to be checked through applications to other educational tests.
{We note that a simple model, such as the one applied in the case study, may be more preferable for the detection of cheating in educational tests, even though more general models are available as discussed in Section~\ref{sec:model}.
This is because the numbers of test takers with preknowledge and compromised items are usually small in an educational test, which makes the effective sample sizes small for estimating parameters related to the outlier classes.
In that case, a more complex model may lead to a high variance in the estimation, which further yields inaccurate classifications. 
}

{Limitations of the proposed method have been discussed in Section~\ref{subsubsec:cheating}. as well as its robustness against model misspecification in Section~\ref{subsec:studyII}.} Another limitation  is that it only models a specific type of cheating, i.e., preknowledge due to item leakage. It does not handle other types of cheating behaviours, such as copying others' answers, electronic transmission of data, hiring stand-ins, and bribing test administrators to correct one's answers. To investigate different types of cheating behaviours, different sources of information are needed and suitable statistical methods remain to be developed. For example, to detect copying behaviour, a statistical model is needed to characterise the similarity between the item responses from two test takers, possibly taking into account their response process information (e.g., response time), seat locations in a test centre, etc. 
We leave these problems for future investigation.

Missing data are widely encountered in educational tests that may be informative for the detection of cheating, though not observed in our real data example. For an educational test with cheating test takers, the missingness of a response likely depends on whether the test taker is cheating and whether the item is compromised.
If many missing responses are observed, then the current framework should be extended by modelling the probabilities of responding. This problem is left for future development, for which ideas may be borrowed from latent variable models for non-ignorable missingness \citep[e.g.,][]{kuha2018latent,o1999symmetric}.

{Besides the applications to cheating detection in educational tests, future research will be conducted to investigate the computation, model evaluation and comparison {in  other areas of application, such as voting behaviours and psychological measurement.} Specifically,  MCMC algorithms for the inference of the proposed model will be further explored. Although our parallel tempering algorithm works well for the current analysis, its performance will be evaluated under more settings, especially  large-scale settings (larger numbers of individuals and items). In addition, other tempering methods, such as simulated tempering, can be explored. Moreover,  goodness-of-fit issues and model selection  will be further studied. In particular, the use of Bayes factors and BIC for comparing the proposed model with several relevant models   will be investigated.} 

\appendix

\bigskip
\noindent
{\bf \LARGE Appendix}

\section{Proof of Proposition~\ref{prop:fdr}}

\begin{proof}
We let $p_{(1)} < \cdots < p_{(n)} \in (0, 1)$ be all the distinct values for $P(\xi_i=1\vert \mathbf Y)$, $i = 1, ..., N$,
where $n$ is less than or equal to $N$ as there might be ties.
We further let $p_{(0)} = 0$ and $p_{(n+1)} = 1$. Then by the form of local FDR in \eqref{eq:fdr}, it is easy to verify that
$\text{fdr}_{\zeta}(Y)$ is a step function of $\zeta$, where $\text{fdr}_{\zeta}(Y)$ is a constant in interval $[p_{(t-1)}, p_{(t)})$, for any $t = 1, ..., n+1$. Therefore, $\text{fdr}_{\zeta}(Y)$ is left continuous in $\zeta$.

We further show that $\text{fdr}_{\zeta}(Y) > \text{fdr}_{\zeta'}(Y)$, when $\zeta \in [p_{(t-1)}, p_{(t)})$ and $\zeta' \in [p_{(t)}, p_{(t+1)})$, for any $t = 1, ..., n$. When $t < n$, we have
$$\text{fdr}_{\zeta'}(Y) = \frac{\sum_{i=1}^N (1-P(\xi_i = 1\vert \mathbf Y)) 1_{\{P(\xi_i = 1\vert \mathbf Y)\geq p_{(t+1)}\}}}{\sum_{i=1}^N 1_{\{P(\xi_i = 1\vert \mathbf Y)\geq p_{(t+1)}\}}},$$
and
$$\text{fdr}_{\zeta}(Y) = \frac{\big(\sum_{i=1}^N (1-P(\xi_i = 1\vert \mathbf Y)) 1_{\{P(\xi_i = 1\vert \mathbf Y)\geq p_{(t+1)}\}}\big) + \big(\sum_{i=1}^N (1-p_{(t)}) 1_{\{P(\xi_i = 1\vert \mathbf Y) =  p_{(t)}\}}\big)}{\sum_{i=1}^N 1_{\{P(\xi_i = 1\vert \mathbf Y)\geq p_{(t+1)}\}} + \sum_{i=1}^N 1_{\{P(\xi_i = 1\vert \mathbf Y) =  p_{(t)}\}}}.$$
As $1-p_{(t)} > 1-P(\xi_i = 1\vert \mathbf Y)$ when $P(\xi_i = 1\vert \mathbf Y)\geq p_{(t+1)}$, $\text{fdr}_{\zeta}(Y) > \text{fdr}_{\zeta'}(Y)$.
When $t = n$, it is easy to see that $\text{fdr}_{\zeta}(\YY) > \text{fdr}_{\zeta'}(\YY)$ as $\text{fdr}_{\zeta'}(\YY) =  0$. This completes the proof for the properties of $\text{fdr}_{\zeta}(Y)$.
The proof for the non-decreasing property of $\text{fnr}_{\zeta}(\YY)$ is similar and thus is omitted here.

By the left-continuity of $\text{fdr}_{\zeta}(\YY)$, we have
$$\text{fdr}_{\zeta^*(\mathbf Y;\rho)}(\YY)\leq \rho.$$
In addition, by the construction of $\zeta^*(\YY;\rho)$, $\zeta' > \zeta^*(\YY;\rho)$ for any $\zeta'\neq \zeta^*(\YY;\rho)$ also satisfying $\text{fdr}_{\zeta'}(\YY)\leq \rho$. Then by the non-decreasing property of $\text{fnr}_{\zeta}(\YY)$, $\text{fnr}_{\zeta'}(\YY) \geq \text{fnr}_{\zeta^*(\mathbf Y;\rho)}(\YY)$.  Therefore, $\zeta^*(\mathbf Y;\rho)$ solves the optimisation problem \eqref{eq:crit}.
\end{proof}

\section{A Parallel Tempering Algorithm}

As mentioned in Section 4.2, the standard MCMC algorithms, such as the Metropolis-Hastings algorithm, suffer from slow-mixing for our problem, due to the presence of many discrete variables and the interactions between these discrete variables in the current problem.

Let $\Xi$ be a generic notation for the parameters and hyperparameters to be sampled. Note that $\Xi = \{\theta_i, \xi_i, \beta_j, \eta_j, \boldsymbol\nu_1, \boldsymbol\nu_2, \delta: i = 1, ..., N, j = 1, ..., J\}$ for the reduced model, and $\Xi = \{\theta_i, \xi_i, \tau_i, \beta_j, \eta_j, \alpha_j, \boldsymbol\nu_1, \boldsymbol\nu_2, \delta, \gamma, \kappa: i = 1, ..., N, j = 1, ..., J\}$ for the full model, respectively. Recall that $\mathbf Z$ is used as the generic notation for data, where $\mathbf Z = \mathbf Y$ and $(\mathbf Y, \mathbf T)$ for the reduced model and the full model, respectively. We use $f(\Xi\vert \mathbf Z)$ as a generic notation for the posterior distribution of interest. The goal is to sample $\Xi$ from the target posterior distribution $f(\Xi\vert \mathbf Z)$.

The algorithm involves sampling $K$ MCMC chains with tempered target distributions. More specifically, let
$0 < \psi_1 < \psi_2 < \ldots < \psi_K = 1$ be a pre-specified sequence of temperature levels. Then the $k$th chain has a target distribution
$f_k(\Xi\vert \mathbf Z) \propto (f(\Xi\vert \mathbf Z))^{(1/\psi_k)}$, where $\propto$ means that the two sides differ by a normalising constant which does not depend on $\Xi$. The target distribution of the $K$th chain is our target posterior distribution. Let $t$ be the current iteration number and $\Xi^{k,t}$ be the current samples from the
$k$th chain. The parallel tempering algorithm   performs the following steps in the $t+1$th iteration.

\begin{enumerate}
    \item For each of the chains, sample $\Xi^{k,t+1}$ given $\Xi^{k,t}$ using a Metropolis-Hastings  within Gibbs sampler, which will be further discussed below.
    \item Randomly sample a pair of adjacent chains, $k$ and $k+1$, and use a Metropolis-Hastings update to decide whether to swap the statuses of  $\Xi^{k,t+1}$ and $\Xi^{k+1,t+1}$.
    That is,  a Bernoulli random variable with success probability
    $$\min\left\{1, \frac{f_k(\Xi^{k+1,t+1}\vert Z) f_{k+1}(\Xi^{k,t+1}\vert Z)}{f_k(\Xi^{k,t+1}\vert Z) f_{k+1}(\Xi^{k+1,t+1}\vert Z)}\right\}$$
    is generated to decide whether to swap or not. If the Bernoulli random variable takes value 1, then we swap the statuses of
  $\Xi^{k,t+1}$ and $\Xi^{k+1,t+1}$ and otherwise, we reject the swap and keep their statuses unchanged.
\end{enumerate}

For simplicity, the MCMC sampling within each chain is conducted by using a  Metropolis-Hastings  within Gibbs sampler. That is, $\Xi$ is split into multiple blocks. Each block is sampled given all the others, using a random-walk Metropolis-Hastings sampler, for which the step size is tuned  following \cite{roberts2001optimal} that is based on the Metropolis-Hastings acceptance rate. For the reduced model, $\Xi$ is split into 10 blocks, including (1) $\theta_i$, $i=1, ..., N$, (2) $\xi_i$, $i=1, ..., N$, (3) $\beta_j$, $j=1, ..., J$, (4) $\eta_j$, $j=1, ..., J$, (5) $\delta$, (6) $\pi_1$, (7) $\sigma_{11}$, (8) $\pi_2$, (9) $\mu_1$, and (10) $\omega_{11}$. For the full model, $\Xi$ is split into 14 blocks, including
(1) $\theta_i$, $i=1, ..., N$, (2) $\tau_i$, $i=1, ..., N$, (3) $\xi_i$, $i=1, ..., N$, (4) $\beta_j$, $j=1, ..., J$, (5) $\alpha_j$,
$j=1, ..., J$,
(6) $\eta_j$, $j=1, ..., J$, (7) $\delta$,
(8) $\gamma$, (9) $\kappa$, (10)
$\pi_1$, (11) $\Sigma$, (12) $\pi_2$, (13) $\boldsymbol\mu$, and (14) $\Omega$.

The specification of the number and levels of the temperatures also needs some tuning. A fine-tuned system tends to have faster mixing. We suggest choosing  the number and levels of the temperatures by following the theoretical guidance given in \cite{atchade2011towards} that is based on the Metropolis-Hastings acceptance rate.

\section{An Additional Simulation Study}

We provide an additional simulation study under settings similar to Study I in Section~\ref{sec:sim}, but with smaller values of $J$ to mimic educational tests with relatively smaller item sizes.

\subsection{Settings}
We consider two settings, with  (1) $N = 2,000, J = 50$, and (2) $N = 4,000, J = 100$. The rest of the simulation setting is exactly the same as that of Study I. These two settings are referred to as settings C.S1 and C.S2, respectively.
For each setting, 100 independent data sets are simulated.





\subsection{Results}
The analysis is conducted using the parallel tempering MCMC algorithm described above.
For each dataset, we run 10,000 iterations, with the first 3,000 iterations as the burn-in.  The results are based on the posterior samples from the last 7,000 iterations.

Our results are given in Tables~\ref{apptab:simclass} through \ref{apptab:simest}. The results are similar to those from Study I. Table~\ref{apptab:simclass} gives the AUC values for the classification of test takers and items. Tables~\ref{apptab:simfdr} and \ref{apptab:simfdr2} evaluate the performance of the local FDR control and local FNR control procedures for the classification of test takers and items, respectively. More specifically, Table~\ref{apptab:simfdr} gives the corresponding FDP values when controlling the local FDR at 1\%, 5\%, and 10\% levels for the test takers. Table~\ref{apptab:simfdr2} gives the
FNP values when controlling the local FNR at 1\%, 5\%, and 10\% levels for the items. Finally, the bias and variance for the posterior mean estimator are given in Table~\ref{apptab:simest}.

\begin{table}
  \centering
  \footnotesize
  \begin{tabular}{ccc|cc|cc|cc}
    \hline
        & \multicolumn{4}{c|}{Test taker}  & \multicolumn{4}{c}{Item} \\
        \cline{2-9}
        & \multicolumn{2}{c|}{C.S1} &\multicolumn{2}{c|}{C.S2} & \multicolumn{2}{c|}{C.S1} &\multicolumn{2}{c}{C.S2} \\
    AUC &  Reduced & Full & Reduced & Full &Reduced & Full & Reduced & Full\\
    \hline
    25\%& 0.934 & 0.936 & 0.956 & 0.962  & 0.937 & 0.944 & 0.954 & 0.961 \\
    50\%& 0.955 & 0.954 & 0.960 & 0.971  & 0.964 & 0.963 & 0.972 & 0.976 \\
    75\%& 0.969 & 0.967 & 0.971 & 0.978  & 0.973 & 0.975 & 0.978 & 0.981 \\
    \hline
  \end{tabular}
  \caption{Overall classification performance based on the posterior means of $\xi_i$ and $\eta_j$.
  For each model, each setting, and each target (test taker/item), we show the 25\%, 50\%, and 75\% quantiles of    the AUCs of the corresponding ROC curves from 100 independent datasets.}\label{apptab:simclass}
\end{table}

\begin{table}
  \centering
  \footnotesize
  \begin{tabular}{cccc|ccc|ccc|ccccc}
    \hline
     & \multicolumn{6}{c|}{C.S1}  & \multicolumn{6}{c}{C.S2} \\
            \cline{2-13}
    & \multicolumn{3}{c|}{Reduced}  & \multicolumn{3}{c|}{Full} & \multicolumn{3}{c|}{Reduced}  & \multicolumn{3}{c}{Full} \\
    FDP & 1\% & 5\% &10\%  & 1\% & 5\% &10\%& 1\% & 5\% &10\%  & 1\% & 5\% &10\% \\
    \hline
    25\%& 0.008 & 0.032 & 0.089  & 0.008 & 0.035 & 0.085  & 0.007 & 0.028 & 0.072 &  0.008 & 0.029 & 0.069\\
    50\%& 0.012 & 0.047 & 0.092  & 0.011 & 0.046 & 0.091  & 0.011 & 0.043 & 0.078 &  0.009 & 0.037 & 0.073\\
    75\%& 0.013 & 0.051 & 0.098  & 0.013 & 0.053 & 0.099  & 0.013 & 0.047 & 0.084 &  0.012 & 0.044 & 0.087\\
    \hline
  \end{tabular}
  \caption{Local FDR control for test takers. For each model, each setting, and each local FDR target (1\%/5\%/10\%), we show the 25\%, 50\%, and 75\% quantiles of the FDPs of the corresponding classifications from 100 independent datasets. }\label{apptab:simfdr}
\end{table}

\begin{table}
  \centering
    \footnotesize
  \begin{tabular}{cccc|ccc|ccc|ccccc}
    \hline
     & \multicolumn{6}{c|}{S1}  & \multicolumn{6}{c}{S2} \\
            \cline{2-13}
    & \multicolumn{3}{c|}{Reduced}  & \multicolumn{3}{c|}{Full} & \multicolumn{3}{c|}{Reduced}  & \multicolumn{3}{c}{Full} \\
    FNP & 1\% & 5\% &10\%  & 1\% & 5\% &10\%& 1\% & 5\% &10\%  & 1\% & 5\% &10\% \\
    \hline
    25\% & 0.009 & 0.033 & 0.063  & 0.007 & 0.032 & 0.061  & 0.007 & 0.029 & 0.059  & 0.006 & 0.031 & 0.062 \\
    50\% & 0.012 & 0.037 & 0.068  & 0.009 & 0.036 & 0.067  & 0.009 & 0.038 & 0.067  & 0.007 & 0.036 & 0.065 \\
    75\% & 0.013 & 0.046 & 0.071  & 0.010 & 0.043 & 0.069  & 0.012 & 0.045 & 0.072  & 0.012 & 0.041 & 0.071 \\
    \hline
  \end{tabular}
  \caption{Local FNR control for items. For each model, each setting, and each local FNR target (1\%/5\%/10\%), we show the 25\%, 50\%, and 75\% quantiles of the FNPs of the corresponding classifications from 100 independent datasets. }\label{apptab:simfdr2}
\end{table}

\begin{table}
\centering
\footnotesize
\begin{tabular}{lcccccc|lcccccccc}
   \hline
    C.S1       & \multicolumn{6}{c|}{Reduced model} &C.S2       & \multicolumn{6}{c}{Reduced model} \\
   \hline
           &$\pi_1$ & $\pi_2$ & $\sigma_{11}$ & $\mu_1$ & $\omega_{11}$ &$\delta$&&$\pi_1$ & $\pi_2$ & $\sigma_{11}$ & $\mu_1$ & $\omega_{11}$ &$\delta$\\
   \hline
  Bias  &  0.11 & -0.02 & -0.05 & -0.11 & 0.17 & 0.22
  &Bias &  0.05 & 0.15 & -0.04 & -0.13 & 0.20 & -0.07 \\
  Variance & 0.16 & 0.09 & 0.32 & 0.43 & 0.17 & 0.30
  &Variance& 0.21 & 0.11 & 0.39 & 0.25 & 0.22 & 0.22\\
   \hline
\end{tabular}
\begin{tabular}{lccccccccccccc}
   \hline
    C.S1       & \multicolumn{12}{c}{Full model}& ~~~~~~~~~ \\
           \hline
           &$\pi_1$ &$\pi_2$ & $\sigma_{11}$ &$\mu_1$ & $\omega_{11}$ &$\delta$ &$\sigma_{22}$ &$\sigma_{12}$ &$\mu_2$ &$\omega_{22}$ &$\omega_{12}$ &$\kappa$ &  \\
           \hline
  Bias
& -0.11 & -0.08 & 0.07 & 0.24 & -0.08 & 0.11 & -0.12 & -0.04 & -0.12 & 0.14 & -0.08 & -0.16 \\
  Variance
& 0.17 & 0.19 & 0.32 & 0.32 & 0.21 & 0.30 & 0.11 & 0.17 & 0.15 & 0.23 & 0.00 & 0.37 \\
     \hline
    C.S2       & \multicolumn{12}{c}{Full model}& ~~~~~~~~~ \\
           \hline
           &$\pi_1$ &$\pi_2$ & $\sigma_{11}$ &$\mu_1$ & $\omega_{11}$ &$\delta$ &$\sigma_{22}$ &$\sigma_{12}$ &$\mu_2$ &$\omega_{22}$ &$\omega_{12}$ &$\kappa$ &  \\
           \hline
  Bias
& -0.04 & 0.06 & 0.08 & -0.15 & -0.17 & -0.15 & -0.04 & -0.11 & -0.15 & 0.11 & -0.04 & -0.11\\
  Variance
& 0.21 & 0.18 & 0.23 & 0.29 & 0.31 & 0.35 & 0.07 & 0.19 & 0.12 & 0.12 & 0.18 & 0.46 \\
  \hline
\end{tabular}
\caption{Accuracy of the posterior mean estimator of
the global parameters. The bias and variance for the posterior mean estimator are calculated based on the 100 replications. }\label{apptab:simest}
\end{table}

\clearpage

\bibliographystyle{apalike}
\bibliography{ref}

\end{document}